# The relation between the ADF and the ionized nebular mass in PNe

Miriam Peña,[1]* Francisco Ruiz-Escobedo[1] and Brenda N. Juárez Torres,[2]
[1] *Universidad Nacional Autónoma de México, Instituto de Astronomía, Apdo. Postal 70264, 04510, Ciudad de México*
[2] *Universidad Nacional Autónoma de México, Facultad de Ciencias, Av. Universidad 3000, 04510, Ciudad de México*

**ABSTRACT**
In this work we analyze the proposed relation between ADFs and ionized masses in planetary nebulae. For this, we have collected from the literature the ADFs and other parameters such as heliocentric distances, H$\beta$ luminosities, logarithmic reddening correction at H$\beta$, c(H$\beta$), electron densities and others and we calculated the ionized mass for a sample of 132 PNe, 27 of which possess a binary central star (14 are close binaries). In addition the distribution of these objects in the Galaxy is analyzed. The ionized masses were calculated considering two different electron densities, the one provided by the [S II] density sensitive lines ratio and the one provided by the [Cl III] lines ratio. No relation was found between ionized masses and ADFs for this sample, although it is confirmed than the PNe with the largest ADFs correspond in general to objects with a close binary central star, although it is important to say that about 20 percent of these objects have an ADF larger than 5 but smaller than 10. Therefore a PN having a close binary central star does not necessarily exhibit an extremely large ADF. We also have searched for possible relations between the ADFs and the stellar atmospheres, divided in H-rich and H-poor stars. No particular relation was found. Interestingly, several PNe with a [WR] H-poor CSPN present an ADF larger than 10, but so far they have not been reported as having a binary central star.

## 1 INTRODUCTION

A very important unresolved problem in the determination of chemical abundances of photo-ionized nebulae is the abundance discrepancy problem which appears when abundances are derived by using collisionally excited lines and recombination lines. It is parameterized by the so called Abundance Discrepancy Factor (ADF) which is defined as the ratio between the abundance derived from recombination lines (RLs) and that derived from collisionally excited lines (CELs). Usually it is measured for the O$^{+2}$/H$^+$ abundance ratio.

$$ADF(O^{+2}) = N(O^{+2})_{RLs}/N(O^{+2})_{CELs}. \quad (1)$$

When available, the ADF has been derived for other ions such as C$^{+2}$, N$^+$ and Ne$^{+2}$. As ADF(O$^{+2}$) is the most commonly measured value in the literature we will use it, and henceforth it will be denoted simply as ADF throughout the text.

The abundance discrepancy problem occurs in H II regions and planetary nebulae (PNe) as well. It was first reported by Wyse (1942). In general, it has been found that ADF values are between 1.5 and 3 in H II regions (Peimbert et al. 2017) and they are larger in PNe, with values between 1.5 and 5, but values larger than 5 have been found in a number of PNe (Corradi et al. 2015; Peimbert et al. 2017). There are a few extreme cases where ADFs larger than several tens have been reported, such as the case of Abell 30 with an ADF of about 600 − 770 in its central filaments (Wesson et al. 2003), Abell 46 with an ADF of about 120 in its central region (Corradi et al. 2015), Abell 58 showing an ADF of 90 (Wesson et al. 2008) and Hf 2-2, with an ADF of 83 (Wesson et al. 2018).

Even ADF values as low as 2 imply a big problem in the chemistry of photo-ionized nebulae because it is not clear which are the real chemical abundances in the ionized medium and this has important repercussions in the analysis of several subjects such as the stellar evolution, the chemical evolution of galaxies and the Universe, etc.

Several solutions have been proposed to explain the ADF problem (see Espíritu & Peimbert 2021 for a complete list). The two most important solutions have been extensively analyzed. The first one is the possible presence of large temperature fluctuations in a chemically homogeneous plasma (Peimbert et al. 2017, and references therein); the fluctuations are larger than those predicted by photo-ionization models, and they have been used mainly to analyze ADFs in H II regions through the parameter $t^2$ that represents the size of the temperature fluctuations (Esteban et al. 2009 and Méndez-Delgado et al. 2023). The second solution corresponds to bi-abundance models where small H-deficient inclusions would be mixed with the H-rich hot plasma (Liu et al. 2000), these inclusions would be cold with $T_e$ of about 1,000 K, and heavy element recombination lines would originate predominantly in this plasma. According to several authors (e. g., Espíritu & Peimbert 2021; Liu 2006; Peimbert & Peimbert 2013; Wesson et al. 2005) ADF values as large as 5 could be explained by the presence of temperature fluctuations in a chemically homogeneous plasma.

However, in several works the possibility of different plasmas of slightly different chemistry and physical conditions, coexisting in the nebulae, cannot be discarded. This possibility has been studied in the works by García-Rojas et al. (2016, 2022), Peña et al. (2017), Richer et al. (2013, 2019, 2022), Ruiz-Escobedo & Peña (2022), Espíritu & Peimbert (2021), and others, and can be studied through kinematic analysis or spatial distribution analysis.

---

* E-mail: miriam@astro.unam.mx

A possible correlation between the ADF and the properties of PNe has been explored in some works of the literature (see, e. g., Wesson et al. 2018, and references therein). In the analysis of ADFs in several PNe, Corradi et al. (2015) and Wesson et al. (2018) found that ADF values seem to be associated to the binarity of the central stars, and in the case of close binary stars the ADFs tend to be large. Corradi et al. (2015) suggest that the ADF value might be related with the ionized mass of the PNe. In this work we aim to explore this suggestion, therefore, we collected data from the literature to derive the ionized masses of PNe and to search for a possible relation between these two parameters. We also explore the possible relation between ADFs and the stellar atmosphere.

This paper is organized as follows. In §2 the analyzed PNe sample is presented and its distribution in the Galaxy is studied. In §3 the ionized masses of PNe are calculated after determining the [S II] and [Cl III] electron densities. In §4 the ADF values and their possible relation with the central star atmosphere are presented. In §5 the possible relation between ADFs and the ionized mass is discussed. Our conclusions are presented in §6.

## 2 THE SAMPLE OF ANALYSED PNE

Our sample consists of 132 Galactic PNe from the list compiled by R. Wesson at https://www.nebulousresearch.org/adfs/, last updated on 9th January 2024. It is the largest compilation of PNe with determined values of ADF. We have also included the recent ADF values determined for IC 4663 (Mohery et al. 2023) and for M 3-27 (Ruiz-Escobedo et al. 2024) that have not yet been included in Wesson's compilation. In Table A1 we list the common names, PN G number, heliocentric distances D adopted from the work by Hernández-Juárez et al. (2024), central star binary, spectral classification of the central star as presented in the catalog by Weidmann et al. (2020), nebular diameters in arcsec, $X_G$, $Y_G$ and $Z_G$ coordinates (coordinates projected on the Galactic plane and height above the Galactic plane, measured relative to the Sun position), and the Galactocentric distance $R_G$. Binary central stars were taken from the compilation by Dr. D. Jones, available at https://www.drdjones.net/bcspn/, last updated on 15th September 2024 (Jones & Boffin 2017; Boffin & Jones 2019).

The values for $X_G$, $Y_G$ and $Z_G$ are given by:

$$X_G = D \cos l \cos b \quad (2)$$

$$Y_G = D \sin l \cos b \quad (3)$$

$$Z_G = D \sin b \quad (4)$$

where the origin of coordinates is the Sun position and $l$ and $b$ are the Galactic longitude and latitude, respectively.

The Galactocentric distance is derived from:

$$R_G = \sqrt{(X_G - X_{GC})^2 + (Y_G - Y_{GC})^2} \quad (5)$$

where $X_{GC}$ = 8 kpc and $Y_{GC}$ = 0 kpc are the coordinates of the Galactic centre (Stanghellini & Haywood 2010).

The Galactic position of the objects in the sample is shown in Fig. 1 where PNe with a close binary central star[1] are shown as blue diamonds. As expected, the PNe are close to the Galactic plane and more concentrated towards the Galactic centre (Acker et al. 1992). A couple of PNe located in the Galactic halo are marked.

---
[1] A close binary star is defined as a binary where the rotation period is ≤

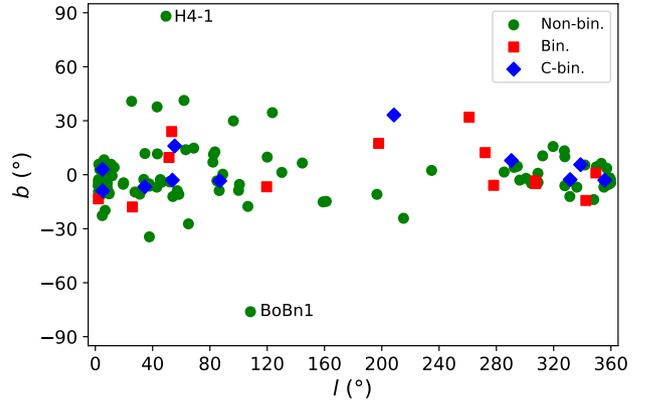

**Figure 1.** The Galactic distribution of the PNe (l, b) in the sample is shown. PNe with distant binary central stars are shown as red squares, PNe with close binary central star are shown as blue diamonds and PNe without a binary central star are shown as green dots. A couple of PN in the galactic halo are marked.

In Fig. 2 the ($X_G$,$Y_G$) distribution of the PNe of the sample is shown. Most of the PNe have $X_G$ and $Y_G$ values lower than 10 kpc. In this figure, the two objects at largest distances are Hen 2-436, which belongs to the Sagittarius spheroidal galaxy (Zijlstra & Walsh 1996), and Sp 4-1, which is a halo PN.

The distribution in $Z_G$ is presented in Fig. 3 where it is verified that most of the PNe are very close to the Galactic plane except those objects identified as belonging to the Galactic halo.

In these figures it is observed that PNe with a binary central star are in the region $-3.7$ kpc $\leq X_G \leq 8.8$ kpc and $-2.6$ kpc $\leq Y_G \leq 5.3$ kpc, while PNe without a reported binary central star are distributed in a larger zone, $-4.4$ kpc $\leq X_G \leq 15$ kpc (Hen 2-436 is not considered) and $-10$ kpc $\leq Y_G \leq 10$ kpc (without considering Sp 4-1). This would be indicating that PNe with a binary star are located nearer the solar vicinity. It is possible that a selection effect is perturbing this graph, because some PNe marked as "not having a binary central star" (shown in green) could possess one not discovered so far.

In Fig. 3, showing the distribution of the sample relative to the Galactic plane, it is observed that most of the objects are very near the Galactic plane, except those belonging to the Galactic halo: H 4-1, DdDm 1, Sp 4-1 and BoBn 1. PNe with a binary central star seem more concentrated but, again, this could be due to a selection effect.

## 3 THE IONIZED MASS

Corradi et al. (2015) discussed a procedure to calculate the ionized mass of Abell 46, Abell 66 and Ou 5 by assuming that there exist two different plasmas in the nebula, a hot, one with low metallicity and emitting mainly the CELs, and a cooler one, with high metallicity and emitting mainly the RLs. The procedure is based on the total flux of H$\beta$, the distance, the electron density and the temperature of the gas determined from CELs and RLs. They concluded that the majority of the H is in the hot gas component and the H$\beta$ emission is mainly

---

1.15 days (Wesson et al. 2018). The periods were obtained from https://www.drdjones.net/bcspn/.

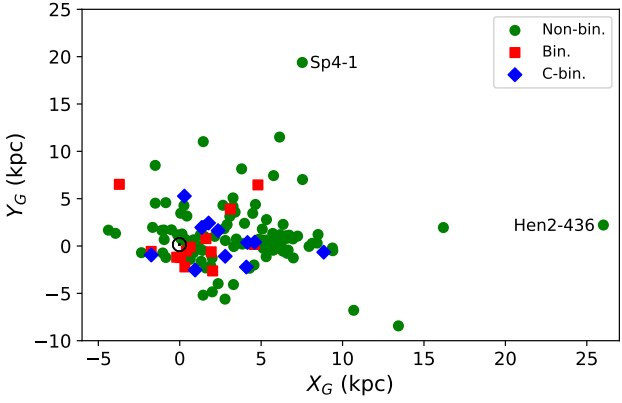

**Figure 2.** The ($X_G$,$Y_G$) distribution of the PNe of the sample is shown. Symbols and colours are the same as in Fig. 1. ⊙ represents the solar position.

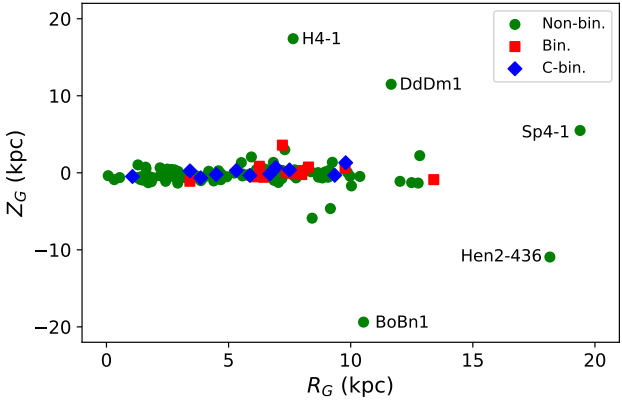

**Figure 3.** The Galactic distribution of PNe in the sample, relative to the Galactic plane, is shown. Most of the objects are very close to the Galactic plane. Several PNe of the Galactic halo are marked. Symbols and colours are the same as in Fig. 1.

produced in the low metallicity zone where the temperature is given by the [O III] lines while the high metallicity zone (with abundances given by RLs) contains a small amount of H and it emits a small amount of H$\beta$. Therefore, following this conclusion, we calculate the ionized mass by assuming that most of the H resides in the hot component and that the nebular ionized mass can be calculated following the expression by Stasińska et al. (2013):

$$M_{ion} = \frac{37.5 \times L_{H\beta}}{n_e} \quad (6)$$

where $M_{ion}$ is the ionized mass in solar units, L$_{H\beta}$ is the total de-reddened luminosity in H$\beta$ in solar units, and $n_e$ is the electron density in particles per cm$^3$. Eq. 6 is equivalent to the expression given by Corradi et al. (2015) and in our case it has been assumed that the electron temperature is $10^4$ K which represents a good average of the $T_e$([O III]) for PNe (see e.g., Henry et al. 2004).

The electron density can be determined with different density sensitive line ratios, being the more used [S II]$\lambda\lambda$6731/6717, [O II]$\lambda\lambda$3729/3726 and [Cl III]$\lambda\lambda$5537/5517. We have compiled from the literature the density sensitive line ratios of [S II] and [Cl III] which are more easily available than the [O II] lines, that are in the ultraviolet and separated by only 2 Å. Thus, electron densities were calculated from [S II] and [Cl III] lines in a homogeneous way by assuming in all cases a $T_e = 10^4$ K and by using the routine *get.TemDen* from PyNeb (Luridiana et al. 2015), version 1.1.18, and the atomic data listed in Table 1. The derived values are listed in Table A2 where we also included the references for the adopted line ratios. These density values were used to derived ionized masses according to Eq. 6.

**Table 1.** Atomic data used for $n_e$ calculations.

| Ion | Transition probabilities | Collisional strenght |
|---|---|---|
| S$^+$ | Podobedova et al. (2009) | Tayal & Zatsarinny (2010) |
| Cl$^{+2}$ | Mendoza (1983) | Butler & Zeippen (1989) |

The luminosity L(H$\beta$) can be derived from the measured flux at H$\beta$ (corrected for reddening using the logarithmic extinction coefficient value, c(H$\beta$), listed in column 2 of Table A3), and the distance $D$ through the next expression:

$$L(H\beta) = 4\pi D^2 F(H\beta) \quad (7)$$

To estimate the total L(H$\beta$) we need to consider if the value of the flux F(H$\beta$) found in the literature is global (including all the nebula) or measured through a slit, usually smaller than the nebular size. In Table A3 we present the H$\beta$ fluxes (from slit and global) obtained from the literature, and the luminosities derived from them.

In Table A4 the derived ionized masses are presented. In different columns it is established if the mass was derived from the total L(H$\beta$) or from a partial value obtained from a slit measurement. To search for the relation between ADFs and ionized masses, only the masses derived from the total L(H$\beta$) were considered due to the masses derived from slit data are usually not including the complete nebula.

## 4 THE VALUES OF ADF

As it was said in §2, the values of ADF for the PNe sample were directly obtained from the compilation by R. Wesson[2], complemented with the two ADF values taken from the works by Mohery et al. (2023); Ruiz-Escobedo et al. (2024), and are listed in Table A4. They correspond to ADFs calculated by different authors (references presented in the same table), according to the Eq. 1 given in the Introduction. A histogram with the ADF values is shown in Fig. 4. In this figure it is observed that the ADF values for PNe without a reported binary central star (shown in blue) are between 2 and 6, except for two objects with ADF of about 40. On the other hand, the values for PNe with a reported binary central star are distributed in a wider rank, up to a value of 770. In the zone of log ADF larger than 1.0 all the objects, except two, correspond to PNe with a binary central star. As a matter of fact, for the 11 PNe with close binary central star shown in Fig. 6, it is found that all of them have ADF value larger than 5, and only 3 objects of this sample present an ADF smaller than 10. It is interesting to notice that an important fraction of PNe with a binary star does not present a extremely large ADF.

---

[2] https://www.nebulousresearch.org/adfs/

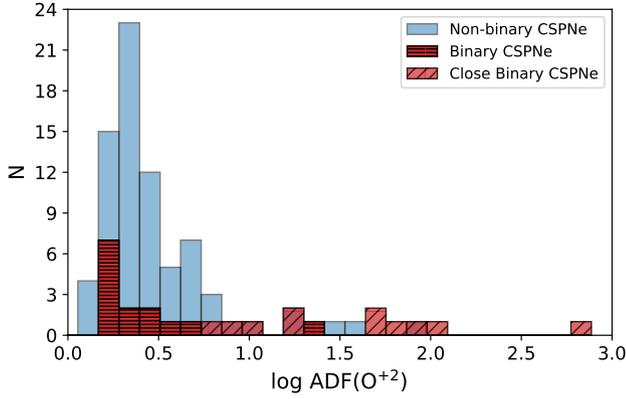

**Figure 4.** The distribution of ADF values. In blue we show the regular PNe and in red, the PNe with a binary central star or a close binary central star.

### 4.1 The ADF and the central star atmosphere

The spectral analysis of central stars (CSPNe) shows that the stellar atmospheres could be classified as H-rich or H-poor. Among the latter ones we find [WR] stars, PG 1159 stars and some H-poor white dwarfs. In the works by García-Rojas et al. (2009, 2013) 15 PNe with a [WR] central star (hence a H-poor central star) were analyzed finding that these objects have low to normal ADF. This is even the case of NGC 5189, which was lately reported as having a binary central star with a period of 4 days (Manick et al. 2015). In this section, we extend the works by García-Rojas et al. (2009, 2013) by analyzing more than a hundred objects possessing different types of central stars: 48 H-rich and 32 H-poor, which include [WR], PG 1159 and other H-poor type stars.

In order to search for possible relations between the ADFs and the central star atmosphere, we use the spectral classification given in the catalog by Weidmann et al. (2020), listed in Table A1. From our original sample of 132 PNe, only 103 PNe posses a classified CSPN in the list by Weidmann et al. (2020). In this catalog, occasionally only the primary star is classified in binary stars and the secondary appears with a quotation mark.

In Fig. 5 the distribution of ADFs grouped by the central star atmosphere is shown. Binary central stars (distant and close) are indicated with different symbols. The upper panel presents the H-rich CSPNe (48 objects distributed in 29 non-binary stars, 9 binaries and 10 close binaries). It is clear that the PNe with ADFs larger than 5 have close binary stars. In the middle panel the objects with H-poor CSPNe (32 objects distributed in 27 non-binary stars, 4 binaries and 1 close binary) show a distribution more concentrated towards lower ADFs although there are five objects with ADF larger than 10 of which only Abell 30, appearing to the right with an ADF = 770 measured in its central knots, has been identified as having a close binary star. It is tempting to suggest that the other 4 objects which are Abell 58, IC 4663, NGC 1501, and NGC 40, all of which have a [WR] star, could posses also a close binary central star. In the lower panel with not classified central stars there are 23 objects distributed in 21 non-binary stars, 1 binary and 1 close binary. PNe included in this group are objects with central stars reported as *wels*, cont., and EL CSPNe in the list by Weidmann et al. (2020). These objects present a distribution similar to the upper panel and the only object with a close binary star is NGC 6337 that possesses an ADF of 18.

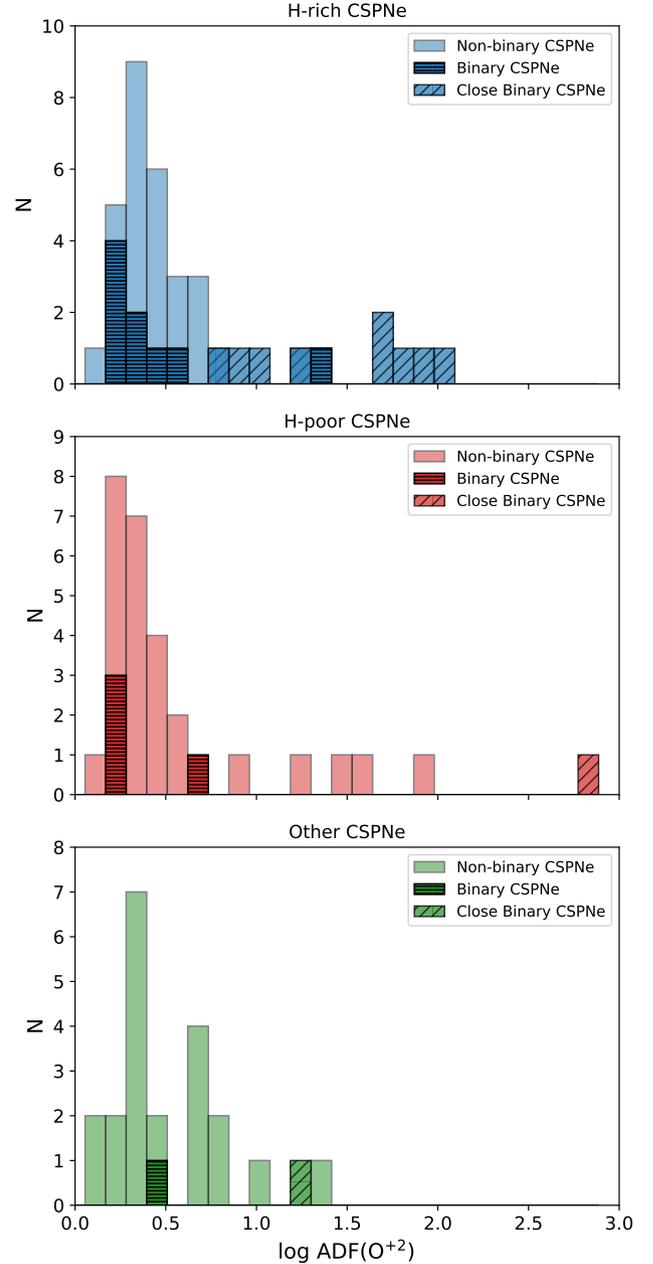

**Figure 5.** The distribution of ADF values grouped by the type of central star atmosphere. In blue the H-rich CSPNe are shown, in red the H-poor CSPNe are indicated, and in green, the other classification by Weidmann et al. (2020) are pointed out.

NGC 6337 is next to M 1-42, a non-binary object with an ADF of 20. It is possible that M 1-42 has a not-reported binary central star.

## 5 THE ADFS AND THE NEBULAR IONIZED MASSES

Fig. 6 shows the relation between the ADF values and the ionized mass derived as described in §3 by using our computed [S II] electron densities, and the global luminosity of H$\beta$. No obvious relation between these two parameters is found, although is it evident that several PNe with a close binary central star have the largest ADFs (in agreement with the results presented previously by Corradi et al. 2015 and other authors). There are also many PNe with a binary star (distant) which present a much smaller ADF as it is also evident in Fig. 5. The PNe with no binary star are distributed all over the diagram. The same is found when the [Cl III] density is used, as it is shown in Fig. 7 where no evident relation between ADFs and ionized masses is found. Unfortunately, there is only one close binary central star object (Hf 2-2) in this diagram with [Cl III] density determined and showing a large ADF.

| $M_{ion}$ | Data number | $R_s$ | $P_v$ |
|---|---|---|---|
| [S II] Slit | 96 | 0.0300 | 0.7713 |
| [Cl III] Slit | 77 | 0.1928 | 0.0929 |
| [S II] Global | 116 | 0.0384 | 0.6810 |
| [Cl III] Global | 90 | 0.1062 | 0.3189 |

**Table 2.** Results from the Spearman statistical test showing the relation between ionized masses and the ADFs.

To check better that there is no relation between these parameters, we run a statistical test. The Spearman correlation coefficient was used. The results are presented in Table 2 where the $R_s$ and $P_v$ values are listed. Both values indicate no correlation between the parameters.

In relation to the ionized mass of the objects, in Figs. 6 and 7 it is clearly observed that PNe with close binary central stars have all type of masses with log $M_{ion}$ from −2 to almost 0 (in solar masses). In this sense, PNe with close binary stars behave equal to PNe with not binary star reported. The PNe with the lowest ionized masses in our sample are Hen 2-86, that has a $M_{ion}$[S II] ∼ $6 \times 10^{-3}$ $M_\odot$, and Vy 2-2, that has a $M_{ion}$[Cl III] ∼ $3.4 \times 10^{-3}$ $M_\odot$. They have not a binary central star, and have reported ADFs of 1.9 and 4.3, respectively.

Santander-García et al. (2022) calculated the ionized and molecular masses of 21 post-common envelope (post-CE) PNe with close binary central star. The molecular masses could be determined only in a couple of objects. They concluded that post-CE PNe arising from single degenerate (SD) systems are as massive as their single-star "regular" counterparts, with a geometrical average value of 0.15 $M_\odot$ while post-CE PNe arising from double degenerate (DD) systems[3] are considerably more massive than SD systems and "regular" PNe, with a geometrical average mass of 0.31 $M_\odot$. Therefore from the work by Santander-García et al. it is evident that the masses of post-CE PNe are not particularly low and they behave equal to PNe with not binary star.

## 6 CONCLUSIONS

In this work, data for 132 PNe with known ADF value were compiled including their heliocentric distances, galactic coordinates, H$\beta$ fluxes, logarithmic extinction coefficients, c(H$\beta$), spectral types of the central stars (available for 103 objects), and nebular sizes. Twenty seven PNe of the sample have been identified as having a binary central star, fourteen of which are close binary central stars. The electron densities were derived by us in a homogeneous way from [S II] and [Cl III] density sensitive lines ratios obtained from the literature and using routines from PyNeb.

---

[3] According to Santander-García et al. (2022), SD are binary systems where one of their components is a post-asymptotic giant branch (post-AGB) star while DD are binary systems where both components are post-AGB stars.

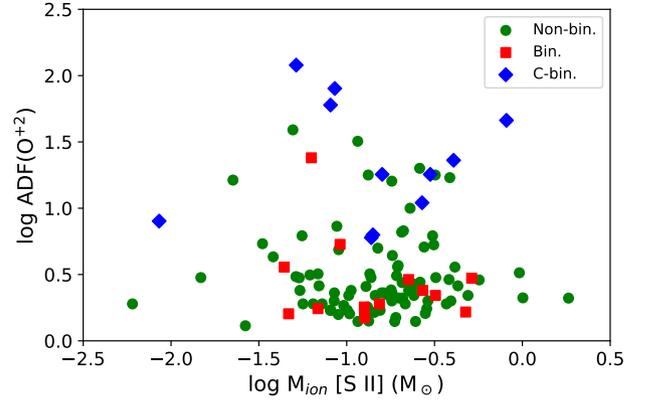

**Figure 6.** Relation between the ionized masses (derived from the [S II] densities and the global H$\beta$ luminosities) vs. the ADFs. PNe with distant binary central stars are shown as red squares, PNe with close binary central stars are shown as blue diamonds and PNe without a binary central stars are shown as green dots.

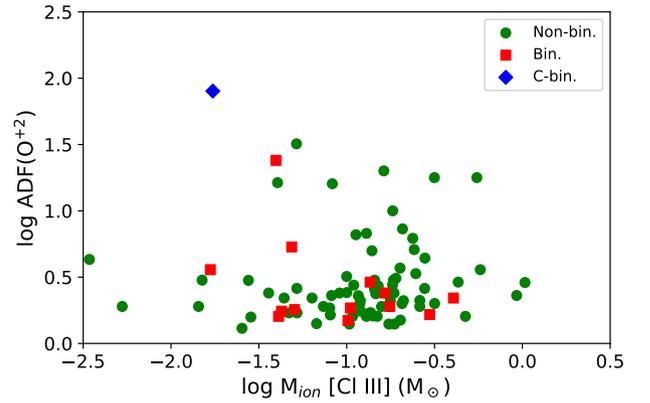

**Figure 7.** Relation between the ionized masses (derived from the [Cl III] densities and the global H$\beta$ luminosities) vs. the ADFs. Symbols and colors are the same as in Fig. 6.

Considering their galactic coordinates, we analyzed the distribution of these objects in the Galaxy calculating their ($X_G$, $Y_G$, $Z_G$) positions, with the Sun at the origin of coordinates and the Galactic centre at 8 kpc in the X direction. From the $X_G$ position of the objects, we found that most of the them are in the direction towards the Galactic centre with values of $X_G$ in the range from −5 kpc (anti-centre direction) to 10 kpc. Regarding the distribution in $Y_G$, most of the objects are within −10 and 10 kpc. The distribution of $Z_G$ values (height above the Galactic plane) shows that most of the objects are near the Galactic plane, having a $Z_G$ value within −3 and 3 kpc with exception of the halo PNe which have larger $Z_G$. The distribution of our objects in the Galaxy is similar to the Galactic distribution of larger samples of PNe (see e.g., Acker et al. 1992).

In §4.1 we have analyzed the possible relation between the ADFs and the stellar type atmospheres. Although one might expect that H-poor central stars can produce nebulae with large ADFs if H-poor material is ejected from the H-poor star, no differences are found in the ADF values for PNe with H-rich and H-poor central stars. In both cases most PNe with large ADFs are those with close binary central stars. Due to this, we are suggesting that Abell 58, IC 4663, NGC 1501, and NGC 40, all of which have a [WC] central star and an ADF larger than 20, might posses a close binary central star.

The main aim of this work was to verify the suggestion by Corradi et al. (2015) about a possible relation between the ADF values and the ionized masses of PNe. Corradi et al. (2015) calculated the ionized masses of three PNe with large ADF values and a close binary star (Abell 43, Abell 65 and Ou 5) which together with results for Hf 2-2 by Liu et al. (2006), led them to the conclusion that PNe with large ADF have low ionized masses. We calculated the ionized masses of a large number of PNe, using the electron densities derived from [S II] and [Cl III] density sensitive line ratios, and by assuming that the electron temperature is $10^4$ K and H$\beta$ luminosity is mainly emitted in the low metallicity zone where the CELs are produced. No attempt was made to calculate the possible contribution to H$\beta$ in the zone where the RLs are mainly produced (a procedure performed by Corradi et al. 2015 and more recently by García-Rojas et al. 2022) due to it is not expected that the H-poor zone emits an important amount of H recombination lines.

No relation was found between the ADFs and the ionized masses, whether visually in the figures or by means of a statistical test (Spearman coefficient). The ionized masses cover a range of log M$_{ion}$ between −2 and 0, in solar masses. It should be noticed that PNe with close binary central star (blue diamonds in Fig. 6) have ionized masses in all this range and the same happens for objects with not binary star. Santander-García et al. (2022) derived a similar result for 21 post-CE PNe. It is important to note that about 20 percent of the PNe in out sample, with a detected close binary star, have an ADF value in the 5 to 10 range, therefore a PN having a close binary central star does not necessarily exhibits an extremely large ADF.

The origin of the ADF in PNe is still an open problem although attempts to correlate it with temperature fluctuations or small H-poor inclusions in the nebulae have been made. Temperature fluctuations seem to work in the case of H II regions (Méndez-Delgado et al. 2023) where the ADFs are lower than 3 but, as mentioned in the Introduction, such a mechanism can not explain ADFs larger than 5.

It has been suggested for some time that there might be different plasmas with different physical conditions and different chemistry coexisting in the nebula and that could be causing the ADFs. The three objects analyzed by Corradi et al. (2015) show such different plasmas with an inner one H-poor and richer in heavy elements and an outer one H-rich and with low metallicity. Also such different plasmas have been found in other PNe, such as the case of NGC 6778, M 1-42, Hf 2-2 and others (García-Rojas et al. 2016, 2022). In addition, different plasmas with different kinematics have been found in some PNe, one emitting mainly the CELs and the other emitting mainly the RLs (Peña et al. 2017; Richer et al. 2013, 2022). When two or more plasma components with different physical conditions are present in the ionized gas of a PN, the relevance of an ADF determination becomes less obvious.

Undoubtedly, much more work is required to disentangle the origin of large ADFs in PNe.

## ACKNOWLEDGEMENTS

We are deeply grateful to Diego B. Hernández-Juárez who helped us with the statistical probes and for providing distance data previous to publication. We thank an anonymous referee whose useful comments helped to improve this article. This work received partial financial support from the grant IN 111423, DGAPA, UNAM.

## DATA AVAILABILITY

The data underlying this article will be shared on reasonable request to the corresponding author.

Table A1: Galactic coordinates, PN G number, heliocentric distance (kpc), binarity (Close-binary, Binary, Non-binary), stellar classification, and diameter of PNe.

| Name | PN G | Distance (kpc) | Binarity | CSPN | Diameter (arcsec) | Diam. ref. | $X_G$ (kpc) | $Y_G$ (kpc) | $Z_G$ (kpc) | $R_G$ (kpc) |
|---|---|---|---|---|---|---|---|---|---|---|
| Abell 30 | 208.5+33.2 | 2.38 | C-bin. | [WC]-PG 1159 + ? | 127.00 | CK71 | −1.748 | −0.951 | 1.306 | 9.794 |
| Abell 46 | 055.4+16.0 | 2.48 | C-bin. | O(H)3 + ? | 63.00 | Cor15 | 1.353 | 1.962 | 0.685 | 6.930 |
| Abell 58 | 037.6−05.1 | 5.68 | Non-bin. | [WC 4-7] | 44.00 | Wes08 | 4.482 | 3.452 | −0.511 | 4.928 |
| Abell 63 | 053.8−03.0 | 3.01 | C-bin. | sdO + ? | 44.90 | Cor15 | 1.771 | 2.428 | −0.159 | 6.685 |
| BoBn 1 | 108.3−76.1 | 19.96 | Non-bin. | — | 2.20 | Ots10 | −1.503 | 4.525 | −19.382 | 10.525 |
| Cn 1-5 | 002.2−09.4 | 4.56 | Non-bin. | [WO 4]pec | 7.00 | WL07 | 4.494 | 0.180 | −0.751 | 3.510 |
| Cn 2-1 | 356.2−04.4 | 6.69 | Non-bin. | Of | 2.40 | WL07 | 6.656 | −0.432 | −0.514 | 1.411 |
| DdDm 1 | 061.9+41.3 | 17.38 | Non-bin. | O(H) | 0.60 | CPT87 | 6.133 | 11.506 | 11.491 | 11.657 |
| Fg 1 | 290.5+07.9 | 2.73 | C-bin. | O(H)3-4 + ? | 16.00 | CKS92 | 0.947 | −2.533 | 0.376 | 7.494 |
| H 1-35 | 355.7−03.4 | 5.74 | Non-bin. | wels? | 2.00 | WL07 | 5.714 | −0.427 | −0.347 | 2.326 |
| H 1-40 | 359.7−02.6 | 7.95 | Non-bin. | — | 3.80 | Mor88 | 7.941 | −0.04 | −0.373 | 0.071 |
| H 1-41 | 356.7−04.8 | 5.37 | Non-bin. | wels | 9.60 | WL07 | 5.343 | −0.302 | −0.449 | 2.675 |
| H 1-42 | 357.2−04.5 | 5.15 | Non-bin. | wels | 5.80 | WL07 | 5.128 | −0.245 | −0.408 | 2.882 |
| H 1-50 | 358.7−05.2 | 9.42 | Non-bin. | — | 10.00 | PK67 | 9.378 | −0.21 | −0.867 | 1.394 |
| H 1-54 | 002.1−04.2 | 8.47 | Non-bin. | — | 4.80 | WL07 | 8.442 | 0.311 | −0.62 | 0.540 |
| H 4-1 | 049.3+88.1 | 17.42 | Non-bin. | — | 2.70 | KM81 | 0.369 | 0.429 | 17.411 | 7.643 |
| Hb 4 | 003.1+02.9 | 3.83 | Non-bin. | [WO 3] | 6.20 | PK67 | 3.819 | 0.212 | 0.195 | 4.186 |
| Hen 2-73 | 296.3−03.0 | 6.27 | Non-bin. | — | 3.30 | Wepg | 2.783 | −5.609 | −0.336 | 7.660 |
| Hen 2-86 | 300.7−02.0 | 4.62 | Non-bin. | [WC 5-6] | 5.60 | CK71 | 2.358 | −3.969 | −0.168 | 6.899 |
| Hen 2-96 | 309.0+00.8 | 5.24 | Non-bin. | — | 4.00 | CK71 | 3.299 | −4.071 | 0.081 | 6.219 |
| Hen 2-118 | 327.5+13.3 | 13.01 | Non-bin. | — | 5.00 | PK67 | 10.685 | −6.789 | 3.000 | 7.301 |
| Hen 2-155 | 338.8+05.6 | 3.01 | C-bin. | O(H)3-5 + ? | 14.50 | CKS92 | 2.793 | −1.083 | 0.298 | 5.319 |
| Hen 2-158 | 327.8−06.1 | 15.95 | Non-bin. | — | 2.00 | CK71 | 13.431 | −8.432 | −1.703 | 10.030 |
| Hen 2-161 | 331.5−02.7 | 4.66 | C-bin. | O(H)3-4 + ? | 10.00 | CKS92 | 4.092 | −2.218 | −0.226 | 4.494 |
| Hen 2-283 | 355.7−03.0 | 8.88 | C-bin. | — | 5.00 | Wepg | 8.843 | −0.651 | −0.479 | 1.065 |
| Hen 2-436 | 004.8−22.7 | 28.31 | Non-bin. | — | 10.00 | PK67 | 26.019 | 2.217 | −10.934 | 18.155 |
| Hen 3-1357 | 331.3−12.1 | 5.00 | Non-bin. | H-rich | 4.00 | Wepg | 4.288 | −2.346 | −1.053 | 4.392 |
| Hf 2-2 | 005.1−08.9 | 4.23 | C-bin. | O(H)3 + ? | 18.60 | CK71 | 4.162 | 0.374 | −0.654 | 3.856 |
| Hu 1-1 | 119.6−06.7 | 7.55 | Bin. | A? + ? | 5.00 | Wes05 | −3.71 | 6.515 | −0.887 | 13.401 |
| Hu 1-2 | 086.5−08.8 | 4.33 | Non-bin. | — | 8.30 | Liu04A | 0.258 | 4.271 | −0.665 | 8.842 |
| Hu 2-1 | 051.4+09.6 | 5.06 | Bin. | WNb + ? | 2.60 | RP22; Wes05 | 3.106 | 3.903 | 0.851 | 6.259 |
| IC 351 | 159.0−15.1 | 4.86 | Non-bin. | O(H)f | 7.00 | Wes05 | −4.38 | 1.676 | −1.273 | 12.493 |
| IC 418 | 215.2−24.2 | 1.40 | Non-bin. | O(H)f | 12.00 | CK71 | −1.043 | −0.736 | −0.576 | 9.073 |
| IC 1747 | 130.2+01.3 | 2.58 | Non-bin. | [WO 4] | 13.00 | Wes05 | −1.667 | 1.968 | 0.063 | 9.865 |
| IC 2003 | 161.2−14.8 | 4.31 | Non-bin. | [WC 3]? | 8.60 | Wes05 | −3.945 | 1.338 | −1.108 | 12.019 |
| IC 3568 | 123.6+34.5 | 2.41 | Non-bin. | O(H)3 | 10.00 | Liu04A | −1.099 | 1.654 | 1.365 | 9.248 |
| IC 4191 | 304.5−04.8 | 2.82 | Non-bin. | — | 5.00 | PK67 | 1.595 | −2.313 | −0.237 | 6.810 |
| IC 4406 | 319.6+15.7 | 1.15 | Non-bin. | [WR] | 35.00 | PK67 | 0.844 | −0.716 | 0.312 | 7.192 |
| IC 4593 | 025.3+40.8 | 3.15 | Non-bin. | O(H)5f | 13.00 | CK71 | 2.154 | 1.020 | 2.060 | 5.934 |
| IC 4663 | 346.2−08.2 | 2.70 | Non-bin. | [WN 3] | 15.67 | Boj21 | −0.748 | −1.953 | −2.505 | 8.964 |
| IC 4699 | 348.0−13.8 | 5.59 | Non-bin. | O(H)3 V((f)) | 5.00 | CK71 | 5.309 | −1.127 | −1.338 | 2.917 |
| IC 4776 | 002.1−13.4 | 4.72 | Bin. | O + ? | 7.50 | CK71 | 4.588 | 0.168 | −1.097 | 3.416 |
| IC 4846 | 027.64−9.6 | 6.10 | Non-bin. | O(H)3-4 f | 2.00 | CK71 | 5.328 | 2.790 | −1.021 | 3.863 |
| IC 4997 | 058.3−10.9 | 5.15 | Non-bin. | wels | 2.50 | Wepg | 2.654 | 4.303 | −0.981 | 6.862 |
| IC 5217 | 100.6−05.4 | 4.68 | Non-bin. | [WC]? | 6.60 | Wes05 | −0.859 | 4.579 | −0.44 | 9.972 |
| K 648 | 065.0−27.3 | 10.12 | Non-bin. | sdO | 1.00 | PK67 | 3.801 | 8.150 | −4.642 | 9.169 |
| M 1-20 | 006.1+08.3 | 7.07 | Non-bin. | wels | 2.50 | Wepg | 6.954 | 0.753 | 1.028 | 1.289 |
| M 1-25 | 004.9+04.9 | 5.60 | Non-bin. | [WC 5-6] | 4.60 | CK71 | 5.559 | 0.480 | 0.482 | 2.488 |
| M 1-29 | 359.1−01.7 | 3.32 | Non-bin. | — | 7.60 | WL07 | 3.318 | −0.052 | −0.099 | 4.682 |
| M 1-30 | 355.9−04.2 | 5.80 | Non-bin. | wels | 5.00 | PK67 | 5.769 | −0.414 | −0.431 | 2.269 |
| M 1-31 | 006.4+02.0 | 5.16 | Non-bin. | wels | — | — | 5.124 | 0.579 | 0.181 | 2.934 |
| M 1-32 | 011.9+04.2 | 3.56 | Non-bin. | [WO 4]pec | 7.60 | CK71 | 3.473 | 0.736 | 0.263 | 4.586 |
| M 1-33 | 013.1+04.1 | 5.88 | Non-bin. | — | 4.80 | CK71 | 5.712 | 1.330 | 0.426 | 2.647 |
| M 1-42 | 002.7−04.8 | 4.37 | Non-bin. | cont. | 9.00 | Liu04B | 4.349 | 0.207 | −0.369 | 3.656 |
| M 1-60 | 019.7−04.5 | 6.77 | Non-bin. | [WC 4] | 10.00 | KO90 | 6.350 | 2.285 | −0.534 | 2.818 |
| M 1-61 | 019.4−05.3 | 5.40 | Non-bin. | wels | 1.80 | WL07 | 5.071 | 1.786 | −0.503 | 3.430 |
| M 1-73 | 051.9−03.8 | 5.32 | Non-bin. | O(H)3.5 If | 5.00 | Wes05 | 3.275 | 4.177 | −0.358 | 6.306 |
| M 1-74 | 052.2−04.0 | 9.43 | Non-bin. | WN b? | 5.00 | Wes05 | 5.763 | 7.435 | −0.659 | 7.764 |
| M 2-4 | 349.8+04.4 | 7.11 | Non-bin. | — | 5.00 | WL07 | 6.976 | −1.255 | 0.553 | 1.620 |
| M 2-6 | 353.3+06.3 | 6.68 | Non-bin. | — | 8.00 | WL07 | 6.595 | −0.769 | 0.734 | 1.602 |
| M 2-23 | 002.2−02.7 | 5.05 | Non-bin. | Of | 8.50 | WL07 | 5.040 | 0.195 | −0.245 | 2.966 |
| M 2-24 | 356.9−05.8 | 9.46 | Non-bin. | — | 6.80 | CK71 | 9.399 | −0.496 | −0.956 | 1.484 |
| Continue | | | | | | | | | | |

**Table A1 – Continued**

| Name | PN G | Distance (kpc) | Binarity | CSPN | Diameter (arcsec) | Diam. ref. | $X_G$ (kpc) | $Y_G$ (kpc) | $Z_G$ (kpc) | $R_G$ (kpc) |
|---|---|---|---|---|---|---|---|---|---|---|
| M 2-27 | 359.9−04.5 | 6.26 | Non-bin. | [WC 4]: | 4.80 | WL07 | 6.240 | −0.005 | −0.501 | 1.760 |
| M 2-31 | 006.0−03.6 | 5.94 | Non-bin. | [WC 4] | 5.10 | Mor88 | 5.895 | 0.624 | −0.375 | 2.195 |
| M 2-33 | 002.0−06.2 | 8.19 | Non-bin. | O(H)5f | 5.80 | WL07 | 8.137 | 0.287 | −0.887 | 0.318 |
| M 2-36 | 003.2−06.1 | 6.17 | Non-bin. | — | 8.00 | Liu01 | 6.124 | 0.351 | −0.665 | 1.909 |
| M 2-39 | 008.1−04.7 | 8.61 | Non-bin. | *wels* | 3.20 | WL07 | 8.494 | 1.215 | −0.713 | 1.312 |
| M 2-42 | 008.2−04.8 | 7.34 | Non-bin. | *wels*: | 3.80 | WL07 | 7.239 | 1.046 | −0.619 | 1.293 |
| M 3-7 | 357.1+03.6 | 5.40 | Non-bin. | *wels* | 5.80 | WL07 | 5.382 | −0.271 | 0.340 | 2.631 |
| M 3-15 | 006.8+04.1 | 5.34 | Non-bin. | [WC 4] | 4.20 | CK71 | 5.288 | 0.631 | 0.387 | 2.784 |
| M 3-21 | 355.1−06.9 | 6.41 | Non-bin. | — | 5.00 | WL07 | 6.340 | −0.543 | −0.776 | 1.747 |
| M 3-27 | 043.3+11.6 | 6.53 | Non-bin. | EL | 1.00 | Wes05 | 4.650 | 4.390 | 1.323 | 5.522 |
| M 3-29 | 004.0−11.1 | 5.73 | Non-bin. | — | 8.20 | WL07 | 5.609 | 0.392 | −1.103 | 2.423 |
| M 3-32 | 009.4−09.8 | 6.70 | Non-bin. | — | 6.00 | WL07 | 6.512 | 1.078 | −1.15 | 1.837 |
| M 3-33 | 009.6−10.6 | 6.94 | Non-bin. | *wels* | 5.00 | WL07 | 6.724 | 1.141 | −1.283 | 1.711 |
| M 3-34 | 031.0−10.8 | 4.73 | Non-bin. | — | 5.60 | Wes05 | 3.979 | 2.396 | −0.894 | 4.681 |
| Me 2-2 | 100.0−08.7 | 8.75 | Non-bin. | Of | 5.00 | Wes05 | −1.506 | 8.516 | −1.333 | 12.763 |
| MPA J1759-3007 | 000.5−03.1 | 7.62 | C-bin. | — | 25.00 | Wepg | 7.608 | 0.066 | −0.423 | 0.398 |
| MyCn 18 | 307.5−04.9 | 3.33 | Bin. | O(H)f + ? | 4.00 | Boj21 | 2.021 | −2.631 | −0.287 | 6.532 |
| NGC 40 | 120.0+09.8 | 1.98 | Non-bin. | [WC 8] | 48.00 | Liu04A | −0.976 | 1.689 | 0.339 | 9.133 |
| NGC 1501 | 144.5+06.5 | 1.18 | Non-bin. | [WO 4] | 52.00 | CKS92 | −0.955 | 0.680 | 0.135 | 8.981 |
| NGC 2022 | 196.6−10.9 | 2.50 | Non-bin. | O(H) | 19.00 | CK71 | −2.351 | −0.705 | −0.474 | 10.375 |
| NGC 2392 | 197.8+17.4 | 1.93 | Bin. | O(H)6f + ? | 46.00 | CKS92 | −1.753 | −0.565 | 0.577 | 9.769 |
| NGC 2440 | 234.8+02.4 | 1.48 | Non-bin. | cont. | 16.00 | PK67 | −0.852 | −1.209 | 0.062 | 8.934 |
| NGC 2867 | 278.1−05.9 | 2.24 | Bin. | [WO 2] + ? | 14.00 | CK71 | 0.316 | −2.206 | −0.231 | 7.994 |
| NGC 3132 | 272.1+12.3 | 1.25 | Bin. | A2 V + ? | 30.00 | CJA87 | 0.045 | −1.22 | 0.268 | 8.048 |
| NGC 3242 | 261.0+32.0 | 1.41 | Bin. | O(H) + ? | 25.00 | CJA87 | −0.186 | −1.181 | 0.748 | 8.271 |
| NGC 3918 | 294.6+04.7 | 1.55 | Non-bin. | O(H)? | 19.00 | CK71; Gar15 | 0.645 | −1.404 | 0.127 | 7.488 |
| NGC 5189 | 307.2−03.4 | 0.68 | Bin. | [WO 1] + ? | 140.00 | CK71 | 0.410 | −0.541 | −0.041 | 7.609 |
| NGC 5307 | 312.3+10.5 | 2.88 | Non-bin. | O(H)3.5 V | 12.50 | CK71 | 1.908 | −2.092 | 0.528 | 6.442 |
| NGC 5315 | 309.1−04.3 | 2.09 | Non-bin. | [WO 4] | 6.00 | CK71 | 1.315 | −1.617 | −0.16 | 6.878 |
| NGC 5882 | 327.8+10.0 | 2.39 | Non-bin. | O(H) f | 14.00 | CK71 | 1.991 | −1.254 | 0.418 | 6.138 |
| NGC 6153 | 341.8+05.4 | 1.48 | Non-bin. | *wels* | 24.00 | CK71 | 1.400 | −0.459 | 0.140 | 6.616 |
| NGC 6210 | 043.1+37.7 | 2.20 | Non-bin. | O(H)3 | 16.20 | Liu04A | 1.270 | 1.189 | 1.347 | 6.834 |
| NGC 6302 | 349.5+01.0 | 0.65 | Bin. | G V + ? | 44.50 | CK71 | 0.639 | −0.118 | 0.012 | 7.362 |
| NGC 6326 | 338.1−08.3 | 3.34 | C-bin. | O(H)5-8 ((fc)) + ? | 12.50 | CK71 | 3.068 | −1.228 | −0.487 | 5.083 |
| NGC 6337 | 349.3−01.1 | 1.82 | C-bin. | *wels* + ? | 51.00 | CJA87 | 1.788 | −0.336 | −0.035 | 6.221 |
| NGC 6369 | 002.4+05.8 | 1.14 | Non-bin. | [WO 3] | 38.00 | CJA87 | 1.133 | 0.048 | 0.116 | 6.867 |
| NGC 6439 | 011.0+05.8 | 6.31 | Non-bin. | — | 5.00 | WL07 | 6.161 | 1.199 | 0.648 | 2.195 |
| NGC 6543 | 096.4+29.9 | 1.37 | Non-bin. | Of-WR(H) | 19.50 | Ack91 | −0.134 | 1.180 | 0.684 | 8.219 |
| NGC 6565 | 003.5−04.6 | 3.92 | Non-bin. | — | 13.60 | WL07 | 3.900 | 0.241 | −0.316 | 4.107 |
| NGC 6567 | 011.7−00.6 | 2.86 | Non-bin. | *wels* | 7.60 | WL07 | 2.800 | 0.582 | −0.032 | 5.232 |
| NGC 6572 | 034.6+11.8 | 1.86 | Non-bin. | Of-WR(H) | 10.80 | Liu04A | 1.498 | 1.034 | 0.382 | 6.584 |
| NGC 6620 | 005.8−06.1 | 6.31 | Non-bin. | cont. | 8.00 | WL07 | 6.241 | 0.643 | −0.675 | 1.873 |
| NGC 6644 | 008.3−07.3 | 4.95 | Non-bin. | *wels* | 2.60 | CKS92 | 4.857 | 0.716 | −0.63 | 3.223 |
| NGC 6720 | 063.1+13.9 | 0.80 | Non-bin. | hgO(H) | 76.00 | Liu04A | 0.350 | 0.693 | 0.193 | 7.681 |
| NGC 6741 | 033.8−02.6 | 3.29 | Non-bin. | — | 8.00 | Liu04A | 2.731 | 1.828 | −0.154 | 5.577 |
| NGC 6778 | 034.5−06.7 | 2.88 | C-bin. | O(H)3-4 + ? | 15.80 | CK71 | 2.355 | 1.624 | −0.337 | 5.874 |
| NGC 6790 | 037.8−06.3 | 3.71 | Non-bin. | WN? | 7.00 | Liu04A | 2.910 | 2.265 | −0.407 | 5.571 |
| NGC 6803 | 046.4−04.1 | 4.95 | Non-bin. | *wels* | 5.50 | Wes05 | 3.402 | 3.578 | −0.356 | 5.826 |
| NGC 6807 | 042.9−06.9 | 10.38 | Non-bin. | Of | 2.00 | Wes05 | 7.541 | 7.022 | −1.251 | 7.037 |
| NGC 6818 | 025.8−17.9 | 1.88 | Bin. | *wels* + ? | 20.00 | CK71 | 1.610 | 0.780 | −0.578 | 6.438 |
| NGC 6826 | 083.5+12.7 | 1.31 | Non-bin. | O(H)3f + ? | 25.00 | Liu04A | 0.143 | 1.269 | 0.290 | 7.959 |
| NGC 6833 | 082.5+11.3 | 11.34 | Non-bin. | Of | 2.00 | Wes05 | 1.445 | 11.024 | 2.230 | 12.826 |
| NGC 6879 | 057.2−08.9 | 6.12 | Non-bin. | O(He)3 f | 5.00 | Wes05 | 3.273 | 5.083 | −0.948 | 6.941 |
| NGC 6884 | 082.1+07.0 | 3.22 | Non-bin. | WN b? | 6.00 | Liu04A | 0.438 | 3.165 | 0.397 | 8.198 |
| NGC 6891 | 054.1−12.1 | 2.83 | Non-bin. | O(H)3 Ib(f*) | 15.00 | Wes05 | 1.619 | 2.244 | −0.594 | 6.764 |
| NGC 7009 | 037.7−34.5 | 1.30 | Non-bin. | O(H) | 28.00 | CKS92 | 0.846 | 0.656 | −0.738 | 7.184 |
| NGC 7026 | 089.0+00.3 | 3.46 | Non-bin. | [WO 3] | 20.00 | Wes05 | 0.060 | 3.459 | 0.022 | 8.661 |
| NGC 7027 | 084.9−03.4 | 0.78 | Non-bin. | cont. | 14.00 | Ack91 | 0.069 | 0.775 | −0.047 | 7.969 |
| NGC 7662 | 106.5−17.6 | 1.87 | Non-bin. | O(H) | 17.00 | Liu04A | −0.508 | 1.709 | −0.565 | 8.678 |
| Ou 5 | 086.9−03.4 | 5.29 | C-bin. | K + ? | 48.00 | Cor15 | 0.285 | 5.273 | −0.321 | 9.345 |
| PB 8 | 292.4+04.1 | 5.25 | Non-bin. | [WN/C] | 5.00 | CKS92 | 1.999 | −4.84 | 0.381 | 7.710 |
| PC 14 | 336.2−06.9 | 5.02 | Non-bin. | [WO 4] | 7.00 | CK71 | 4.562 | −2.004 | −0.611 | 3.980 |
| Pe 1-1 | 285.4+01.5 | 5.39 | Non-bin. | [WC 5] | 3.00 | CK71 | 1.435 | −5.193 | 0.141 | 8.371 |
| Continue | | | | | | | | | | |

Table A1 – Continued

| Name | PN G | Distance (kpc) | Binarity | CSPN | Diameter (arcsec) | Diam. ref. | $X_G$ (kpc) | $Y_G$ (kpc) | $Z_G$ (kpc) | $R_G$ (kpc) |
|---|---|---|---|---|---|---|---|---|---|---|
| Pe 1-9 | 005.0+03.0 | 4.63 | C-bin. | O(H) + ? | 12.40 | CK71 | 4.606 | 0.405 | 0.247 | 3.419 |
| Sp 3 | 342.5−14.3 | 2.09 | Bin. | O3 + ? | 36.00 | PK67 | 1.932 | −0.609 | −0.517 | 6.099 |
| Sp 4-1 | 068.7+14.8 | 21.51 | Non-bin. | O? | — | — | 7.527 | 19.386 | 5.495 | 19.392 |
| Vy 1-2 | 053.3+24.0 | 8.81 | Bin. | [WR]/*wels* + ? | 4.60 | RP22; Wes05 | 4.802 | 6.455 | 3.590 | 7.203 |
| Vy 2-1 | 007.0−06.8 | 6.69 | Non-bin. | *wels* | 7.00 | WL07 | 6.592 | 0.814 | −0.796 | 1.626 |
| Vy 2-2 | 045.4−02.7 | 4.39 | Non-bin. | B[e] | 3.10 | RP22 | 3.074 | 3.127 | −0.207 | 5.835 |
| Wray 16-423 | 006.8−19.8 | 17.34 | Non-bin. | — | 2.60 | Ots15B | 16.192 | 1.951 | −5.891 | 8.421 |

Distances from Hernández-Juárez et al. (2024).
CSPN classifications from Weidmann et al. (2020).
References: Ack91:Acker et al. (1991), Boj21: Bojičić et al. (2021), CK71: Cahn & Kaler (1971), CJA87: Chu et al. (1987), CPT87: Clegg et al. (1987), Cor15: Corradi et al. (2015), Gar15: García-Rojas et al. (2015), KO90: Kapusta & Olive (1990), KM81 Kohoutek & Martin (1981), Liu01: Liu et al. (2001), Liu04A: Liu et al. (2004a), Liu04B: Liu et al. (2004b), Mor88: Moreno et al. (1988), Ots10: Otsuka et al. (2010), Ots15B: Otsuka (2015), PK67: Perek & Kohoutek (1967), RP22: Ruiz-Escobedo & Peña (2022), WL07: Wang & Liu (2007), Wes05: Wesson et al. (2005), Wes08: Wesson et al. (2008), Wepg: Wesson (2024).

Table A2: Line ratios from the literature and calculated electron densities from [S II] and [Cl III] lines with PyNeb. A $T_e = 10^4$ K was assumed for all calculations.

| Name | [S II] 6731/6716 | [Cl III] 5537/5517 | Reference | $n_e$ ([S II]) (cm$^{-3}$) | $n_e$ ([Cl III]) (cm$^{-3}$) |
|---|---|---|---|---|---|
| Abell 46 | 0.90 | — | Cor15 | 313 | — |
| Abell 58 | 1.40 | — | Wes08 | 1628 | — |
| Abell 63 | 1.08 | — | Cor15 | 666 | — |
| BoBn 1 | 1.51 | — | Ots10 | 2140 | — |
| Cn 1-5 (a) | 1.72 | 1.18 | WL07 | 3611 | 3200 |
| Cn 1-5 (b)* | 1.76 | 1.22 | Gar12 | 4020 | 3504 |
| Cn 2-1 | 1.72 | 1.18 | WL07 | 3611 | 3200 |
| DdDm 1 (a)* | 1.69 | 0.40 | Ots09 | 3344 | — |
| DdDm 1 (b) | 2.71 | — | Wes05 | — | — |
| Fg 1 | 0.89 | — | Wes18 | 297 | — |
| H 1-35 | 2.17 | 3.14 | WL07 | 20378 | 35502 |
| H 1-40 | 1.99 | 1.35 | Gar18 | 8257 | 4530 |
| H 1-41 | 1.23 | 0.89 | WL07 | 1048 | 1192 |
| H 1-42 | 1.85 | 1.45 | WL07 | 5185 | 5381 |
| H 1-50 (a)* | 1.92 | 1.97 | Gar18 | 6455 | 10780 |
| H 1-50 (b) | 1.85 | 1.79 | WL07 | 5185 | 8693 |
| H 1-54 | 2.00 | 2.16 | WL07 | 8575 | 13259 |
| H 4-1 | 1.14 | 1.00 | Ots13 | 806 | 1913 |
| Hb 4 | 1.90 | 1.51 | Gar12 | 6058 | 5914 |
| Hen 2-118 | 1.94 | 1.97 | WL07 | 6909 | 10780 |
| Hen 2-155 | 1.22 | — | Jon15 | 1018 | — |
| Hen 2-158 | 1.62 | 1.47 | Gar18 | 2806 | 5550 |
| Hen 2-161 | 1.27 | — | Jon15 | 1168 | — |
| Hen 2-283 | 1.71 | — | Wes18 | 3525 | — |
| Hen 2-436 | 2.25 | 2.81 | Ots11 | 45827 | 25560 |
| Hen 2-73 | 1.93 | 1.94 | Gar18 | 6683 | 10412 |
| Hen 2-86 | 2.13 | 2.46 | Gar12 | 15798 | 18072 |
| Hen 2-96 | 2.04 | 2.56 | Gar18 | 10087 | 19981 |
| Hen 3-1357 (a) | 2.06 | 2.05 | Ots17 | 11044 | 11785 |
| Hen 3-1357 (b) | 1.93 | — | Pen22 | 6683 | — |
| Hen 3-1357 (c)* | 1.92 | 2.20 | Pen22 | 6455 | 13853 |
| Hf 2-2 (b)* | 0.92 | 0.97 | McN16 | 347 | 1716 |
| Hu 1-1 | 1.25 | 0.66 | Wes05 | 1107 | — |
| Hu 1-2 | 1.67 | 1.27 | Liu04A | 3167 | 3889 |
| Hu 2-1 (a)* | 2.31 | 2.08 | RP22 | — | 12184 |
| Hu 2-1 (b) | 2.35 | — | Wes05 | — | — |
| IC 351 | 1.75 | — | Wes05 | 3912 | — |
| IC 418 | 2.12 | 1.96 | Sha04 | 14917 | 10651 |
| IC 1747 | 1.71 | — | Wes05 | 3525 | — |
| IC 2003 | 1.70 | 0.93 | Wes05 | 3431 | 1450 |
| IC 3568 | 1.29 | — | Liu04A | 1232 | — |
| IC 4191 (a)* | 1.99 | 2.10 | Tsa03 | 8257 | 12444 |
| IC 4191 (b) | 1.08 | 1.92 | Tsa03 | 666 | 10156 |
| IC 4406 | 1.13 | 1.22 | Tsa03 | 781 | 3504 |
| IC 4593 | 1.23 | 0.82 | Rob05 | 1048 | 752 |
| IC 4663 | 1.52 | 0.65 | Moh23 | 2182 | — |
| IC 4699 | 1.52 | 0.65 | WL07 | 2182 | — |
| IC 4776 | 2.07 | 2.49 | Sow17 | 11556 | 18626 |
| IC 4846 (a)* | 1.83 | 1.36 | WL07 | 4907 | 4620 |
| IC 4846 (b) | 2.07 | — | Wes05 | 11556 | — |
| IC 4997 | 2.17 | — | — | 20378 | — |
| IC 5217 | 0.49 | — | Wes05 | — | — |
| K 648 | 1.53 | 1.33 | Ots15A | 2242 | 4362 |
| M 1-20 | 1.93 | 1.80 | WL07 | 6683 | 8798 |
| M 1-25 | 1.99 | 0.06 | Gar12 | 8257 | — |
| M 1-29 | 1.59 | 1.35 | WL07 | 2600 | 4530 |
| M 1-30 | 1.90 | 1.63 | Gar12 | 6058 | 7046 |
| M 1-31 | 1.80 | 2.42 | Gar18 | 4496 | 17350 |
| M 1-32 | 2.08 | 2.06 | Gar12 | 12110 | 11911 |
| M 1-33 | 1.79 | 1.55 | Gar18 | 4375 | 6273 |
| M 1-42 (a)* | 1.21 | 0.95 | Liu01 | 990 | 1584 |
| M 1-42 (b) | 1.24 | 0.94 | McN16 | 1077 | 1515 |

**Table A2 – Continued**

| Name | [S II] 6731/6716 | [Cl III] 5537/5517 | Reference | $n_e$ ([S II]) (cm$^{-3}$) | $n_e$ ([Cl III]) (cm$^{-3}$) |
|---|---|---|---|---|---|
| M 1-60 | 1.91 | 2.05 | Gar18 | 6244 | 11785 |
| M 1-61 (a)* | 2.17 | 2.36 | Gar12 | 20378 | 16332 |
| M 1-61 (b) | 2.08 | 2.45 | WL07 | 12110 | 17909 |
| M 1-73 | 1.80 | — | Wes05 | 4496 | — |
| M 1-74 | 2.09 | — | Wes05 | 12729 | — |
| M 2-4 | 1.82 | 1.60 | WL07 | 4761 | 6744 |
| M 2-6 | 1.86 | 0.85 | WL07 | 5345 | 939 |
| M 2-23 | 2.05 | 2.05 | WL07 | 10539 | 11785 |
| M 2-24 | 1.32 | — | Zha03 | 1333 | — |
| M 2-27 | 1.87 | 2.13 | WL07 | 5517 | 12865 |
| M 2-31 | 1.81 | 1.75 | Gar18 | 4620 | 8257 |
| M 2-33 | 1.33 | 0.55 | WL07 | 1367 | — |
| M 2-36 (a) | 1.66 | 1.41 | Liu01 | 3096 | 5027 |
| M 2-36 (b)* | 1.61 | 1.34 | Esp21 | 2737 | 4448 |
| M 2-39 | 1.72 | 0.93 | WL07 | 3611 | 1450 |
| M 2-42 | 1.58 | 1.15 | WL07 | 2538 | 2972 |
| M 3-7 | 1.77 | 1.11 | WL07 | 4119 | 2688 |
| M 3-15 | 1.88 | 1.72 | Gar12 | 5720 | 7951 |
| M 3-21 | 1.98 | 1.96 | WL07 | 7963 | 10651 |
| M 3-27 (b)* | 1.27 | — | RPB24 | 1168 | — |
| M 3-29 | 1.08 | 0.58 | WL07 | 666 | — |
| M 3-32 | 1.50 | 0.84 | WL07 | 2089 | 877 |
| M 3-33 | 1.15 | 0.94 | WL07 | 831 | 1515 |
| Me 2-2 | 1.10 | — | Wes05 | 710 | — |
| MPA J1759-3007 | 1.10 | — | Wes18 | 710 | — |
| MyCn 18 | 1.76 | 1.91 | Tsa03 | 4020 | 10034 |
| NGC 40 | 1.37 | 0.84 | Liu04A | 1512 | 877 |
| NGC 1501 | 1.10 | 0.95 | Erc04 | 710 | 1584 |
| NGC 2022 | 1.12 | 0.96 | Tsa03 | 757 | 1650 |
| NGC 2392 (a) | 1.56 | 0.94 | Zha12 | 2415 | 1515 |
| NGC 2392 (b)* | 1.08 | 0.87 | Zha12 | 666 | 1067 |
| NGC 2440 | 1.94 | — | Tsa03 | 6909 | — |
| NGC 2867 (a)* | 1.63 | 1.23 | Gar09 | 2870 | 3573 |
| NGC 2867 (b) | 1.49 | 1.24 | Gar09 | 2033 | 3650 |
| NGC 3132 | 0.98 | 0.82 | Tsa03 | 458 | 752 |
| NGC 3242 | 1.41 | 0.91 | Tsa03 | 1670 | 1319 |
| NGC 3918 (a) | 1.74 | 1.43 | Tsa03 | 3813 | 5197 |
| NGC 3918 (b)* | 1.82 | 1.49 | Gar15 | 4761 | 5720 |
| NGC 5189 | 1.21 | 0.88 | Gar12 | 990 | 1129 |
| NGC 5307 | 1.78 | 0.94 | Rui03 | 4245 | 1515 |
| NGC 5315 (a)* | 2.01 | 2.90 | Ma17 | 8865 | 27900 |
| NGC 5315 (b) | 1.89 | 2.70 | Tsa03 | 5860 | 22926 |
| NGC 5882 | 1.72 | 1.21 | Tsa03 | 3611 | 3420 |
| NGC 6153 (a)* | 1.66 | 1.27 | Liu00 | 3096 | 3889 |
| NGC 6153 (b) | 1.68 | 0.44 | McN16 | 3264 | — |
| NGC 6210 (b) | 1.66 | 1.17 | Liu04A | 3096 | 3124 |
| NGC 6302 | 1.96 | 2.54 | Tsa03 | 7440 | 19593 |
| NGC 6326 | 1.12 | — | Wes18 | 757 | — |
| NGC 6337 | 0.91 | — | Wes18 | 330 | — |
| NGC 6369 | 1.67 | 1.21 | Gar12 | 3167 | 3420 |
| NGC 6439 | 1.74 | 1.40 | WL07 | 3813 | 4945 |
| NGC 6543S | 1.85 | 1.40 | Wes04 | 5185 | 4945 |
| NGC 6565 | 1.34 | 0.95 | WL07 | 1401 | 1584 |
| NGC 6567 | 1.83 | 2.17 | WL07 | 4907 | 13441 |
| NGC 6572 | 2.11 | 1.25 | Liu04A | 14042 | 3733 |
| NGC 6620 | 1.49 | 1.12 | WL07 | 2033 | 2754 |
| NGC 6720 | 0.96 | 0.78 | Liu04A | 420 | 509 |
| NGC 6741 | 1.74 | 1.37 | Liu04A | 3813 | 4697 |
| NGC 6778 | 1.06 | — | Jon16 | 621 | — |
| NGC 6790 | 2.19 | 2.46 | Liu04A | 24280 | 18072 |
| NGC 6803 | 1.88 | 1.81 | Wes05 | 5720 | 8905 |
| NGC 6807 | 2.03 | — | Wes05 | 9669 | — |
| NGC 6818 | 1.33 | 1.05 | Tsa03 | 1367 | 2263 |
| NGC 6826 | 1.37 | 0.91 | Liu04A | 1512 | 1319 |

**Table A2 –** *Continued*

| Name | [S II] 6731/6716 | [Cl III] 5537/5517 | Reference | $n_e$ ([S II]) (cm$^{-3}$) | $n_e$ ([Cl III]) (cm$^{-3}$) |
|---|---|---|---|---|---|
| NGC 6884 | 1.88 | 1.42 | Liu04A | 5720 | 5127 |
| NGC 7009 | 1.66 | 1.20 | Fan11 | 3096 | 3344 |
| NGC 7026 | 2.33 | 1.82 | Wes05 | — | 9014 |
| NGC 7027 | 2.25 | 3.42 | Zha05 | 45827 | 47879 |
| NGC 7662 | 1.50 | 1.00 | Liu04A | 2089 | 1913 |
| Ou 5 | 0.80 | — | Cor15 | 154 | — |
| PB 8 | 1.51 | 1.00 | Gar09 | 2140 | 1913 |
| PC 14 | 1.66 | 1.16 | Gar12 | 3096 | 3045 |
| Pe 1-1 | 2.11 | 2.75 | Gar12 | 14042 | 24098 |
| Pe 1-9 | 0.96 | — | Wes18 | 420 | — |
| Sp 3 | 1.10 | 0.88 | Mis19 | 710 | 1129 |
| Vy 1-2 (a)* | 1.88 | 1.97 | RP22 | 5720 | 10780 |
| Vy 1-2 (b) | 1.22 | — | Wes05 | 1018 | — |
| Vy 2-1 | 1.61 | 1.43 | WL07 | 2737 | 5197 |
| Vy 2-2 (a)* | 1.95 | 3.83 | RP22 | 7121 | 79237 |
| Vy 2-2 (b) | 3.02 | — | Wes05 | — | — |
| Wray 16-423 | 1.67 | 1.48 | Ots15B | 3167 | 5643 |

* Adopted values for ionized mass calculations.

References: Cor15: Corradi et al. (2015), Erc04: Ercolano et al. (2004),
Esp21: Espíritu & Peimbert (2021), Fan11: Fang & Liu (2011), Gar09: García-Rojas et al. (2009),
Gar12: García-Rojas et al. (2012), Gar15: García-Rojas et al. (2015),
Gar18: García-Rojas et al. (2018), Jon15: Jones et al. (2015), Jon16: Jones et al. (2016),
Liu00: Liu et al. (2000), Liu04A: Liu et al. (2004a), Liu06: Liu et al. (2006),
Ma17: Madonna et al. (2017), McN16: McNabb et al. (2016), Mis19: Miszalski et al. (2019),
Moh23: Mohery et al. (2023), Ots09: Otsuka et al. (2009), Ots10: Otsuka et al. (2010),
Ots11: Otsuka et al. (2011), Ots13: Otsuka & Tajitsu (2013), Ots15A: Otsuka et al. (2015),
Ots15B: Otsuka (2015), Ots17: Otsuka et al. (2017), Pen22: Peña et al. (2022),
Rob05: Robertson-Tessi & Garnett (2005), Rui03: Ruiz et al. (2003),
RP22: Ruiz-Escobedo & Peña (2022), RPB24: Ruiz-Escobedo et al. (2024),
Sha04: Sharpee et al. (2004), Sim22: Simpson et al. (2022), Sow17: Sowicka et al. (2017),
Tsa03: Tsamis et al. (2003), WL07: Wang & Liu (2007), Wes03: Wesson et al. (2003),
Wes04: Wesson & Liu (2004), Wes05: Wesson et al. (2005), Wes18: Wesson et al. (2018),
Zha05: Zhang et al. (2005), Zha12: Zhang et al. (2012).

Table A3: c(H$\beta$), sizes of employed slits for H$\beta$ Flux, total fluxes and L(H$\beta$)

| Name | c(H$\beta$) | Ref. c(H$\beta$) | Slit size (arcsec) | F(H$\beta$) Slit (erg cm$^{-2}$ s$^{-1}$) | Ref. F(H$\beta$) Slit | L(H$\beta$) Slit (L$_\odot$) | F(H$\beta$) Global (erg cm$^{-2}$ s$^{-1}$) | Ref. F(H$\beta$) Global | L(H$\beta$) Global (L$_\odot$) |
|---|---|---|---|---|---|---|---|---|---|
| Abell 30 (J1)* | 1.02 | Wes03 | 0.82 | 1.24E−16 | Wes03 | 2.20E−4 | 6.46E−13 | Kal83b | 1.15 |
| Abell 30 (J3) | 0.64 | Wes03 | 0.82 | 1.49E−16 | Wes03 | 1.10E−4 | 6.46E−13 | Kal83b | 0.48 |
| Abell 30 (J4) | 1.26 | Sim22 | 0.82 | 1.37E−16 | Sim22 | 4.22E−4 | 6.46E−13 | Kal83b | 1.99 |
| Abell 46 | 0.22 | Cor15 | 1.00 | 3.53E−14 | Cor15 | 0.01 | 1.41E−12 | Kal83b | 0.43 |
| Abell 58 | 1.04 | Wes08 | 1.00 | 3.80E−17 | Wes08 | 4.02E−4 | — | — | — |
| Abell 63 | 0.55 | Cor15 | 1.00 | 1.80E−14 | Cor15 | 0.02 | 1.58E−13 | Ack91 | 0.15 |
| BoBn 1 | 0.09 | Ots10 | 1.20 | 2.95E−13 | Ots10 | 4.33 | 2.00E−13 | Ack91 | 2.93 |
| Cn 1-5 (a) | 0.49 | WL07 | 2.00 | 3.98E−12 | WL07 | 7.65 | 6.17E−12 | Ack91 | 11.86 |
| Cn 1-5 (b)* | 0.56 | Gar12 | 1.00 | 9.72E−13 | Gar12 | 2.20 | 6.17E−12 | Ack91 | 13.94 |
| Cn 2-1 | 1.07 | WL07 | 2.00 | 2.34E−12 | WL07 | 36.81 | 2.34E−12 | Web83 | 36.81 |
| DdDm 1 (a)* | 0.06 | Hen08 | 0.60 | 1.86E−12 | Ots09 | 19.30 | 2.69E−12 | Ack91 | 27.91 |
| DdDm 1 (b) | 0.14 | Wes05 | 1.00 | 2.69E−12 | Wes05 | 33.56 | 2.69E−12 | Ack91 | 33.56 |
| Fg 1 | 0.52 | Wes18 | — | — | — | — | 8.71E−12 | Per71 | 6.43 |
| H 1-35 | 1.51 | WL07 | 2.00 | 3.16E−12 | WL07 | 100.80 | 3.55E−12 | Ack91 | 113.24 |
| H 1-40 | 2.41 | Gar18 | 1.30 | 1.41E−13 | Gar18 | 68.53 | 2.57E−13 | Ack91 | 124.91 |
| H 1-41 | 0.65 | WL07 | 2.00 | 1.26E−12 | WL07 | 4.86 | 2.00E−12 | Ack91 | 7.71 |
| H 1-42 | 0.87 | WL07 | 2.00 | 2.09E−12 | WL07 | 12.29 | 2.00E−12 | Ack91 | 11.76 |
| H 1-50 (a)* | 0.88 | Gar18 | 1.30 | 1.79E−12 | Gar18 | 36.05 | 2.09E−12 | Ack91 | 42.09 |
| H 1-50 (b) | 0.68 | WL07 | 2.00 | 1.99E−12 | WL07 | 25.29 | 2.09E−12 | Ack91 | 26.56 |
| H 1-54 | 1.54 | WL07 | 2.00 | 1.35E−12 | WL07 | 100.47 | 1.32E−12 | Web83 | 98.24 |
| H 4-1 | 0.10 | Ots13 | 0.60 | 3.86E−13 | Ots13 | 4.41 | 5.37E−13 | Ack91 | 6.14 |
| Hb 4 | 1.81 | Gar12 | 1.00 | 1.34E−13 | Gar12 | 3.80 | 1.12E−12 | Sha89 | 31.74 |
| Hen 2-73 | 1.39 | Gar18 | 1.30 | 9.11E−13 | Gar09 | 26.30 | 1.12E−12 | Ack91 | 32.34 |
| Hen 2-86 | 2.10 | Gar12 | 1.00 | 2.96E−13 | Gar12 | 23.80 | 3.16E−14 | Ack91 | 2.54 |
| Hen 2-96 | 2.01 | Gar18 | 1.00 | 3.35E−13 | Gar18 | 28.16 | 4.27E−13 | Ack91 | 35.89 |
| Hen 2-118 | 0.17 | WL07 | 2.00 | 2.14E−12 | WL07 | 16.03 | 2.00E−12 | Sha89 | 14.98 |
| Hen 2-155 | 0.74 | Jon15 | — | — | — | — | 2.57E−12 | Sha89 | 3.83 |
| Hen 2-158 | 0.56 | Gar18 | 1.30 | 1.01E−12 | Gar18 | 27.91 | 7.59E−13 | Web69 | 20.98 |
| Hen 2-161 | 1.21 | Jon15 | — | — | — | — | 7.94E−13 | Web69 | 8.37 |
| Hen 2-283 | 1.48 | Wes18 | — | — | — | — | 1.82E−13 | Ack91 | 12.97 |
| Hen 2-436 | 0.23 | Ots11 | — | — | — | — | 6.76E−13 | Web83 | 27.53 |
| Hen 3-1357 (a) | 0.16 | Ots17 | 2.00 | 9.84E−12 | Ots17 | 10.64 | — | — | — |
| Hen 3-1357 (b) | 0.15 | Pen22 | 2.00 | 9.84E−12 | Ots17 | 10.40 | — | — | — |
| Hen 3-1357 (c)* | 0.22 | Pen22 | 2.00 | 9.84E−12 | Ots17 | 12.21 | — | — | — |
| Hf 2-2 (a) | 0.47 | Liu06 | — | — | — | — | 5.01E−13 | Ack91 | 0.79 |
| Hf 2-2 (b)* | 0.47 | Liu06 | — | — | — | — | 5.01E−13 | Ack91 | 0.79 |
| Hf 2-2 (c) | 0.33 | Wes18 | — | — | — | — | 5.01E−13 | Ack91 | 0.57 |
| Hu 1-1 | 0.55 | Wes05 | 1.00 | 2.51E−12 | Wes05 | 15.19 | 2.51E−12 | Bar78 | 15.19 |
| Hu 1-2 | 0.51 | Liu04A | 1.00 | 6.17E−12 | Liu04A | 11.20 | 6.17E−12 | Kal76 | 11.20 |
| Hu 2-1 (a)* | 0.45 | RP22 | 1.00 | 1.58E−11 | Wes05 | 34.11 | 1.58E−11 | KM81 | 34.11 |
| Hu 2-1 (b) | 0.78 | Wes05 | 1.00 | 1.58E−11 | Wes05 | 72.93 | 1.58E−11 | KM81 | 72.93 |
| IC 351 | 0.38 | Wes05 | 1.00 | 3.80E−12 | Wes05 | 6.44 | 3.80E−12 | Bar78 | 6.44 |
| IC 418 | 0.34 | Sha04 | — | — | — | — | 2.63E−10 | Sha89 | 33.74 |
| IC 1747 | 1.00 | Wes05 | 1.00 | 3.24E−12 | Wes05 | 6.45 | 3.24E−12 | ODe63 | 6.45 |
| IC 2003 | 0.35 | Wes05 | 1.00 | 6.46E−12 | Wes05 | 8.04 | 6.46E−12 | Bar78 | 8.04 |
| IC 3568 | 0.26 | Liu04A | 1.00 | 1.51E−11 | Liu04A | 4.77 | 1.51E−11 | Kal76 | 4.77 |
| IC 4191 (a)* | 0.70 | Tsa03 | 2.00 | 1.02E−11 | Tsa03 | 12.16 | 1.00E−11 | Sha89 | 11.92 |
| IC 4191 (b) | 0.70 | Tsa03 | 2.00 | 1.02E−11 | Tsa03 | 12.16 | 1.00E−11 | Sha89 | 11.92 |
| IC 4406 | 0.27 | Tsa03 | 2.00 | 1.78E−11 | Tsa03 | 1.31 | 1.82E−11 | Web83 | 1.34 |
| IC 4593 | 0.17 | Rob05 | 2.50 | 2.63E−11 | Rob05 | 11.55 | 2.63E−11 | Web83 | 11.55 |
| IC 4663 | 0.54 | Moh23 | — | — | — | — | 3.80E−12 | Moh23 | 2.87 |
| IC 4699 | 0.23 | WL07 | 2.00 | 2.00E−12 | WL07 | 3.18 | 2.04E−12 | Sha89 | 3.24 |
| IC 4776 | 0.22 | Sow17 | — | — | — | — | 1.91E−11 | Web83 | 21.13 |
| IC 4846 (a)* | 0.69 | WL07 | 2.00 | 5.01E−12 | WL07 | 27.32 | 4.57E−12 | Sha89 | 24.92 |
| IC 4846 (b) | 0.70 | Wes05 | 1.00 | 4.57E−12 | Wes05 | 25.50 | 4.57E−12 | Sha89 | 25.50 |
| IC 4997 | 0.32 | RP22 | — | — | — | — | 2.95E−11 | Kal78 | 48.91 |
| IC 5217 | 0.50 | Wes05 | 1.00 | 6.76E−12 | Wes05 | 14.01 | 6.76E−12 | Bar78 | 14.01 |
| K 648 | 0.12 | Ots15A | 1.20 | 7.83E−13 | Ots15A | 3.16 | 7.94E−13 | Haw78 | 3.21 |
| M 1-20 | 1.40 | WL07 | 2.00 | 1.14E−12 | WL07 | 42.82 | 1.17E−12 | Sha89 | 43.95 |
| M 1-25 | 1.41 | Gar12 | 1.00 | 5.01E−13 | Gar12 | 12.08 | 1.20E−12 | Ack91 | 28.94 |
| M 1-29 | 2.03 | WL07 | 2.00 | 3.89E−13 | WL07 | 13.75 | 6.31E−13 | Ack91 | 22.30 |
| M 1-30 | 1.00 | Gar12 | 1.00 | 5.20E−13 | Gar12 | 5.23 | 1.70E−12 | Ack91 | 17.11 |
| M 1-31 | 2.08 | Gar18 | 1.30 | 3.84E−13 | Gar18 | 36.78 | 1.26E−12 | Ack91 | 120.67 |

Continue

Table A3 – Continued

| Name | c(Hβ) | Ref. c(Hβ) | Slit size (arcsec) | F(Hβ) Slit (erg cm$^{-2}$ s$^{-1}$) | Ref. F(Hβ) Slit | L(Hβ) Slit (L$_\odot$) | F(Hβ) Global (erg cm$^{-2}$ s$^{-1}$) | Ref. F(Hβ) Global | L(Hβ) Global (L$_\odot$) |
|---|---|---|---|---|---|---|---|---|---|
| M 1-32 | 1.30 | Gar12 | 1.00 | 2.01E−13 | Gar12 | 1.52 | 6.31E−13 | Ack91 | 4.77 |
| M 1-33 | 1.56 | Gar18 | 1.30 | 8.65E−13 | Gar18 | 32.49 | 8.13E−13 | Ack91 | 30.53 |
| M 1-42 (a)* | 0.70 | Liu04B | 1.00 | 2.34E−12 | Liu04B | 6.70 | 2.40E−12 | Web83 | 6.87 |
| M 1-42 (b) | 0.70 | Liu04B | 1.00 | 2.34E−12 | Liu04B | 6.70 | 2.40E−12 | Web83 | 6.87 |
| M 1-60 | 1.68 | Gar18 | 1.30 | 6.25E−13 | Gar18 | 41.02 | 5.25E−13 | Ack91 | 34.46 |
| M 1-61 (a)* | 1.24 | Gar12 | 1.00 | 2.45E−12 | Gar12 | 37.14 | 3.72E−12 | Ack91 | 56.40 |
| M 1-61 (b) | 0.92 | WL07 | 2.00 | 3.47E−12 | WL07 | 25.18 | 3.72E−12 | Ack91 | 26.99 |
| M 1-73 | 1.14 | Wes05 | 1.00 | 2.00E−12 | Wes05 | 23.38 | 2.00E−12 | Ack91 | 23.38 |
| M 1-74 | 1.12 | Wes05 | 1.00 | 1.78E−12 | Wes05 | 62.43 | 1.78E−12 | Bar78 | 62.43 |
| M 2-4 | 1.33 | WL07 | 2.00 | 1.15E−12 | WL07 | 37.19 | 1.45E−12 | Ack91 | 46.89 |
| M 2-6 | 1.14 | WL07 | 2.00 | 5.89E−13 | WL07 | 10.85 | 1.26E−12 | Ack91 | 23.22 |
| M 2-23 | 1.20 | WL07 | 2.00 | 2.63E−12 | WL07 | 31.80 | 2.69E−12 | Web83 | 32.53 |
| M 2-24 | 0.80 | Zha03 | 2.00 | 7.94E−13 | Zha03 | 13.41 | 8.13E−13 | Ack91 | 13.73 |
| M 2-27 | 1.31 | WL07 | 2.00 | 5.89E−13 | WL07 | 14.10 | 6.31E−13 | Ack91 | 15.10 |
| M 2-31 | 1.43 | Gar18 | 1.30 | 7.02E−13 | Gar18 | 19.95 | 7.76E−13 | Via85 | 22.05 |
| M 2-33 | 0.55 | WL07 | 2.00 | 1.41E−12 | WL07 | 10.04 | 2.51E−12 | Ack91 | 17.87 |
| M 2-36 (a) | 0.27 | Liu04B | 2.00 | 3.54E−12 | Liu01 | 7.51 | 6.31E−12 | Ack91 | 13.38 |
| M 2-36 (b)* | 0.33 | Esp21 | 2.00 | 3.54E−12 | Liu01 | 8.62 | 6.31E−12 | Ack91 | 15.37 |
| M 2-39 | 0.61 | WL07 | 2.00 | 8.51E−13 | WL07 | 7.69 | 7.41E−13 | Ack91 | 6.70 |
| M 2-42 | 1.06 | WL07 | 2.00 | 7.94E−13 | WL07 | 14.69 | 7.59E−13 | Ack91 | 14.05 |
| M 3-7 | 1.65 | WL07 | 2.00 | 4.17E−13 | WL07 | 16.25 | 5.13E−13 | Web83 | 19.99 |
| M 3-15 | 2.09 | Gar12 | 1.00 | 9.01E−13 | Gar12 | 94.57 | 3.55E−13 | Ack91 | 37.26 |
| M 3-21 | 0.50 | WL07 | 2.00 | 4.07E−12 | WL07 | 15.82 | 3.80E−12 | Sha89 | 14.77 |
| M 3-27 (a) | — | — | 1.00 | 1.48E−12 | Wes05 | — | 1.48E−12 | Bar78 | — |
| M 3-27 (b)* | 0.63 | RPB24 | 1.00 | — | — | — | 5.50E−12 | KM81 | 29.93 |
| M 3-29 | 0.24 | WL07 | 2.00 | 1.66E−12 | WL07 | 2.83 | 2.51E−12 | Ack91 | 4.28 |
| M 3-32 | 0.64 | WL07 | 2.00 | 1.41E−12 | WL07 | 8.27 | 1.26E−12 | Ack91 | 7.39 |
| M 3-33 | 0.50 | WL07 | 2.00 | 1.17E−12 | WL07 | 5.33 | 1.00E−12 | Ack91 | 4.56 |
| M 3-34 | 0.58 | Wes05 | 1.00 | 1.58E−12 | Wes05 | 4.02 | 1.58E−12 | Ack91 | 4.02 |
| Me 2-2 | 0.34 | Wes05 | 1.00 | 6.92E−12 | Wes05 | 34.68 | 6.92E−12 | Bar78 | 34.68 |
| MPA J1759-3007 | 1.46 | Wes18 | — | — | — | — | — | — | — |
| MyCn 18 | 0.74 | Tsa03 | 2.00 | 6.17E−12 | Tsa03 | 11.25 | 7.41E−12 | Web69 | 13.51 |
| NGC 40 | 0.70 | Liu04A | 1.00 | 4.27E−11 | Liu04A | 25.10 | 2.19E−11 | Lil55 | 12.87 |
| NGC 1501 | 1.00 | Erc04 | — | — | — | — | 5.25E−12 | Col61 | 2.19 |
| NGC 2022 | 0.42 | Tsa03 | 2.00 | 7.41E−12 | Tsa03 | 3.64 | 7.41E−12 | KM81 | 3.64 |
| NGC 2392 (a) | 0.27 | Zha12 | — | — | — | — | 4.07E−11 | Sha85 | 8.45 |
| NGC 2392 (b)* | 0.27 | Zha12 | — | — | — | — | 4.07E−11 | Sha85 | 8.45 |
| NGC 2440 | 0.47 | Tsa03 | 2.00 | 3.16E−11 | Tsa03 | 6.11 | 3.16E−11 | Sha89 | 6.11 |
| NGC 2867 (a)* | 0.39 | Gar09 | — | — | — | — | 2.63E−11 | Sha89 | 9.69 |
| NGC 2867 (b) | 0.43 | Gar09 | — | — | — | — | 2.63E−11 | Sha89 | 10.63 |
| NGC 3132 | 0.30 | Tsa03 | 2.00 | 3.55E−11 | Tsa03 | 3.31 | 3.55E−11 | Web69 | 3.31 |
| NGC 3242 | 0.17 | Tsa03 | 2.00 | 1.62E−10 | Tsa03 | 14.25 | 1.62E−10 | Sha89 | 14.25 |
| NGC 3918 (a) | 0.40 | Tsa03 | 2.00 | 9.12E−11 | Tsa03 | 16.47 | 9.12E−11 | KM81 | 16.47 |
| NGC 3918 (b)* | 0.27 | Gar15 | 1.00 | 5.54E−11 | Gar15 | 0.74 | 9.12E−11 | KM81 | 12.21 |
| NGC 5189 | 0.47 | Gar12 | 1.00 | 1.17E−13 | Gar09 | 4.78E−3 | 3.02E−11 | Via85 | 1.23 |
| NGC 5307 | 0.59 | Rui03 | 3.00 | 1.18E−12 | Rui03 | 1.14 | 6.61E−12 | Web83 | 6.38 |
| NGC 5315 (a)* | 0.63 | Ma17 | — | — | — | — | 3.80E−11 | KM81 | 21.18 |
| NGC 5315 (b) | 0.55 | Tsa03 | 2.00 | 3.80E−11 | Tsa03 | 17.62 | 3.80E−11 | KM81 | 17.62 |
| NGC 5882 | 0.42 | Tsa03 | 2.00 | 4.17E−11 | Tsa03 | 18.74 | 4.27E−11 | KM81 | 19.19 |
| NGC 6153 (a)* | 1.30 | Liu00 | — | — | — | — | 1.45E−11 | Web83 | 18.96 |
| NGC 6153 (b) | 1.32 | McN16 | — | — | — | — | 1.45E−11 | Web83 | 19.85 |
| NGC 6210 (a) | 0.53 | Rob05 | 2.50 | 8.13E−11 | Rob05 | 18.05 | 8.13E−11 | KM81 | 18.05 |
| NGC 6210 (b)* | 0.13 | Liu04A | 1.00 | 8.13E−11 | Liu04A | 15.88 | 8.13E−11 | KM81 | 15.88 |
| NGC 6302 | 1.39 | Tsa03 | 2.00 | 2.82E−11 | Tsa03 | 8.75 | 2.82E−11 | Per71 | 8.75 |
| NGC 6326 | 0.47 | Wes18 | — | — | — | — | 8.32E−12 | Sha89 | 8.20 |
| NGC 6337 | 0.50 | Wes18 | — | — | — | — | 4.47E−12 | Per71 | 1.40 |
| NGC 6369 | 1.93 | Gar12 | 1.00 | 6.72E−14 | Gar12 | 0.22 | 4.79E−12 | Per71 | 15.85 |
| NGC 6439 | 1.10 | WL07 | 2.00 | 1.86E−12 | WL07 | 27.89 | 2.09E−12 | Ack91 | 31.34 |
| NGC 6543 (a) | 0.08 | Rob05 | 2.50 | 2.45E−10 | Rob05 | 16.54 | 2.69E−10 | Wes04 | 18.16 |
| NGC 6543 (b) | 0.08 | Rob05 | 2.50 | 2.45E−10 | Rob05 | 16.54 | 2.69E−10 | Wes04 | 18.16 |
| NGC 6543S (c)* | 0.10 | Wes04 | — | — | — | — | 2.69E−10 | Wes04 | 19.02 |
| NGC 6565 | 0.32 | WL07 | 2.00 | 5.62E−12 | WL07 | 5.40 | 6.03E−12 | Sha89 | 5.79 |
| NGC 6567 | 0.90 | WL07 | 2.00 | 1.15E−11 | WL07 | 22.35 | 1.17E−11 | Sha89 | 22.74 |
| NGC 6572 (a) | 1.22 | Rob05 | 2.50 | 1.51E−10 | Rob05 | 259.38 | 1.51E−10 | Sha89 | 259.38 |

Continue

**Table A3 – Continued**

| Name | c(Hβ) | Ref. c(Hβ) | Slit size (arcsec) | F(Hβ) Slit (erg cm$^{-2}$ s$^{-1}$) | Ref. F(Hβ) Slit | L(Hβ) Slit (L$_\odot$) | F(Hβ) Global (erg cm$^{-2}$ s$^{-1}$) | Ref. F(Hβ) Global | L(Hβ) Global (L$_\odot$) |
|---|---|---|---|---|---|---|---|---|---|
| NGC 6572 (b) | 1.22 | Rob05 | 2.50 | 1.51E−10 | Rob05 | 259.38 | 1.51E−10 | Sha89 | 259.38 |
| NGC 6572 (c) | 1.22 | Rob05 | 2.50 | 1.51E−10 | Rob05 | 259.38 | 1.51E−10 | Sha89 | 259.38 |
| NGC 6572 (d)* | 0.48 | Liu04A | 1.00 | 1.51E−10 | Liu04A | 47.20 | 1.51E−10 | Sha89 | 47.20 |
| NGC 6620 | 0.52 | WL07 | 2.00 | 1.86E−12 | WL07 | 7.34 | 1.86E−12 | Web83 | 7.34 |
| NGC 6644 | — | — | — | — | — | — | 1.02E−11 | Sha89 | — |
| NGC 6720 (a)* | 0.20 | Liu04A | 1.00 | 8.32E−11 | Liu04A | 2.52 | 8.32E−11 | Col61 | 2.52 |
| NGC 6720 (b) | — | — | — | — | — | — | 8.32E−11 | Col61 | — |
| NGC 6741 | 1.15 | Liu04A | 1.00 | 4.57E−12 | Liu04A | 20.90 | 4.79E−12 | Web83 | 21.91 |
| NGC 6778 | 0.46 | Jon16 | 0.70 | 6.17E−12 | Jon16 | 4.42 | 6.92E−12 | Kal83b | 4.95 |
| NGC 6790 (a) | 0.45 | Rob05 | 2.50 | 1.26E−11 | Rob05 | 14.62 | 1.26E−11 | Kal83b | 14.62 |
| NGC 6790 (b)* | 1.10 | Liu04A | 1.00 | 1.26E−11 | Liu04A | 65.32 | 1.26E−11 | Kal83b | 65.32 |
| NGC 6803 | 0.87 | Wes05 | 1.00 | 6.61E−12 | Wes05 | 35.92 | 6.61E−12 | Col61 | 35.92 |
| NGC 6807 | 0.64 | Wes05 | 1.00 | 3.31E−12 | Wes05 | 46.58 | 3.31E−12 | Kal80 | 46.58 |
| NGC 6818 | 0.37 | Tsa03 | 2.00 | 3.31E−11 | Tsa03 | 8.21 | 3.31E−11 | Web83 | 8.21 |
| NGC 6826 | 0.06 | Liu04A | 1.00 | 1.10E−10 | Liu04A | 6.48 | 1.05E−10 | Kal78 | 6.19 |
| NGC 6833 | 0.00 | Wes05 | 1.00 | 5.62E−12 | Wes05 | 21.62 | 5.62E−12 | Ack91 | 21.62 |
| NGC 6879 | 0.40 | Wes05 | 1.00 | 2.63E−12 | Wes05 | 7.40 | 2.63E−12 | Web83 | 7.40 |
| NGC 6884 | 1.00 | Liu04A | 1.00 | 7.76E−12 | Liu04A | 24.07 | 7.76E−12 | Col61 | 24.07 |
| NGC 6891 | 0.29 | Wes05 | 1.00 | 2.24E−11 | Wes05 | 10.47 | 2.24E−11 | Web83 | 10.47 |
| NGC 7009 (a) | 0.20 | Liu95 | — | — | — | — | 1.66E−10 | Web83 | 13.30 |
| NGC 7009 (b)* | 0.17 | Fan11 | — | — | — | — | 1.66E−10 | Web83 | 12.41 |
| NGC 7026 | 1.12 | Wes05 | 1.00 | 1.26E−11 | Wes05 | 59.49 | 1.26E−11 | Kal85 | 59.49 |
| NGC 7027 (a) | 0.92 | Rob05 | 2.50 | 7.59E−11 | Rob05 | 11.49 | 7.59E−11 | Sha82 | 11.49 |
| NGC 7027 (b)* | 1.37 | Zha05 | — | — | — | — | 7.59E−11 | Sha82 | 32.39 |
| NGC 7662 | 0.18 | Liu04A | 1.00 | 1.02E−10 | Liu04A | 16.15 | 1.02E−10 | Kal78 | 16.15 |
| Ou 5 | 0.94 | Cor15 | 1.00 | 1.37E−14 | Cor15 | 0.10 | — | — | — |
| PB 8 | 0.36 | Gar09 | — | — | — | — | 3.89E−12 | Sha89 | 7.35 |
| PC 14 | 0.63 | Gar12 | 1.00 | 5.05E−13 | Gar12 | 1.62 | 1.86E−12 | Sha89 | 5.98 |
| Pe 1-1 | 1.80 | Gar12 | 1.00 | 2.57E−13 | Gar12 | 14.09 | 5.50E−13 | Ack91 | 30.16 |
| Pe 1-9 | 1.35 | Wes18 | — | — | — | — | 6.31E−14 | Ack91 | 0.91 |
| Sp 3 | 0.06 | Mis19 | — | — | — | — | 7.94E−12 | Ack91 | 1.19 |
| Sp 4-1 | 0.00 | Wes05 | 1.00 | 1.45E−12 | Wes05 | 20.07 | 1.45E−12 | Ack91 | 20.07 |
| Vy 1-2 (a)* | 0.31 | RP22 | 1.00 | 2.95E−12 | Wes05 | 13.99 | 2.95E−12 | Kal80 | 13.99 |
| Vy 1-2 (b) | 0.14 | Wes05 | 1.00 | 2.95E−12 | Wes05 | 9.46 | 2.95E−12 | Kal80 | 9.46 |
| Vy 2-1 | 0.83 | WL07 | 2.00 | 2.69E−12 | WL07 | 24.35 | 3.16E−12 | Ack91 | 28.61 |
| Vy 2-2 (a)* | 0.66 | RP22 | 1.00 | 2.75E−12 | Wes05 | 7.25 | 2.75E−12 | Sha89 | 7.25 |
| Vy 2-2 (b) | 1.65 | Wes05 | 1.00 | 2.75E−12 | Wes05 | 70.83 | 2.75E−12 | Sha89 | 70.83 |
| Wray 16-423 | 0.13 | Ots15B | 1.20 | 1.53E−12 | Ots15B | 18.57 | 1.00E−12 | Ack91 | 12.14 |

* Adopted values for ionized mass calculations.

References: Ack91: Acker et al. (1991), Bar78: Barker (1978), Col61: Collins et al. (1961), Cor15: Corradi et al. (2015), Erc04: Ercolano et al. (2004), Esp21: Espíritu & Peimbert (2021), Fan11: Fang & Liu (2011), Gar09: García-Rojas et al. (2009), Gar12: García-Rojas et al. (2012), Gar15: García-Rojas et al. (2015), Gar18: García-Rojas et al. (2018), Haw78: Hawley & Miller (1978), Hen08: Henry et al. (2008), Jon15: Jones et al. (2015), Jon16: Jones et al. (2016), Kal76: Kaler (1976), Kal78: Kaler (1978), Kal80: Kaler (1980), Kal83b: Kaler (1983), Kal85: Kaler & Lutz (1985), KM81: Kohoutek & Martin (1981), Lil55: Liller (1955), Liu95: Liu et al. (1995), Liu00: Liu et al. (2000), Liu04A: Liu et al. (2004a), Liu04B: Liu et al. (2004b), Liu06: Liu et al. (2006), Ma17: Madonna et al. (2017), Mis19: Miszalski et al. (2019), Moh23: Mohery et al. (2023), ODe63: O'Dell (1963), Ots10: Otsuka et al. (2010), Ots11: Otsuka et al. (2011), Ots15A: Otsuka et al. (2015), Ots15B: Otsuka (2015), Ots17: Otsuka et al. (2017), Pen22: Peña et al. (2022), Per71: Perek (1971), Rob05: Robertson-Tessi & Garnett (2005), Rui03: Ruiz et al. (2003), RP22: Ruiz-Escobedo & Peña (2022), RPB24: Ruiz-Escobedo et al. (2024), Sha04: Sharpee et al. (2004), Sha82: Shaw & Kaler (1982), Sha85: Shaw & Kaler (1985), Sha89: Shaw & Kaler (1989), Sim22: Simpson et al. (2022), Sow17: Sowicka et al. (2017), Tsa03: Tsamis et al. (2003), Via85: Viadana & de Freitas Pacheco (1985), WL07: Wang & Liu (2007), Web69: Webster (1969), Web83: Webster (1983), Wes03: Wesson et al. (2003), Wes04: Wesson & Liu (2004), Wes05: Wesson et al. (2005), Wes18: Wesson et al. (2018), Zha03: Zhang & Liu (2003), Zha05: Zhang et al. (2005), Zha12: Zhang et al. (2012).

Table A4: ADFs from the list by R. Wesson, ionized masses derived from the densities of [S II] and [Cl III] and the fluxes with slit or global, in solar units, are presented.

| Name | ADF | Ref. ADF | $M_{ion}$([S II]) Slit ($M_\odot$) | $M_{ion}$([Cl III]) Slit ($M_\odot$) | $M_{ion}$([S II]) Global ($M_\odot$) | $M_{ion}$([Cl III]) Global ($M_\odot$) |
|---|---|---|---|---|---|---|
| Abell 30 (J1)* | 770.00 | Wes03 | — | — | — | — |
| Abell 30 (J3) | 600.00 | Wes03 | — | — | — | — |
| Abell 30 (J4) | 22.00 | Sim22 | — | — | — | — |
| Abell 46 | 120.00 | Cor15 | 1.29E−3 | — | 5.16E−2 | — |
| Abell 58 | 90.00 | Wes08 | 9.26E−6 | — | — | — |
| Abell 63 | 8.00 | Cor15 | 9.75E−4 | — | 8.56E−3 | — |
| BoBn 1 | 3.05 | Ots10 | 7.58E−2 | — | 5.14E−2 | — |
| Cn 1-5 (a) | 1.90 | WL07 | 7.95E−2 | 8.97E−2 | 1.23E−1 | 1.39E−1 |
| Cn 1-5 (b)* | 1.60 | Gar12 | 2.05E−2 | 2.35E−2 | 1.30E−1 | 1.49E−1 |
| Cn 2-1 | 2.90 | WL07 | 3.82E−1 | 4.31E−1 | 3.82E−1 | 4.31E−1 |
| DdDm 1 (a)* | 5.30 | Ots09 | 2.16E−1 | — | 3.13E−1 | — |
| DdDm 1 (b) | 11.80 | Wes05 | — | — | — | — |
| Fg 1 | 46.00 | Wes18 | — | — | 8.12E−1 | — |
| H 1-35 | 2.10 | WL07 | 1.85E−1 | 1.06E−1 | 2.08E−1 | 1.20E−1 |
| H 1-40 | 2.88 | Gar18 | 3.11E−1 | 5.67E−1 | 5.67E−1 | 1.03 |
| H 1-41 | 5.10 | WL07 | 1.74E−1 | 1.53E−1 | 2.76E−1 | 2.42E−1 |
| H 1-42 | 2.30 | WL07 | 8.89E−2 | 8.57E−2 | 8.51E−2 | 8.20E−2 |
| H 1-50 (a)* | 2.37 | Gar18 | 2.09E−1 | 1.25E−1 | 2.45E−1 | 1.46E−1 |
| H 1-50 (b) | 2.90 | WL07 | 1.83E−1 | 1.09E−1 | 1.92E−1 | 1.15E−1 |
| H 1-54 | 2.60 | WL07 | 4.39E−1 | 2.84E−1 | 4.30E−1 | 2.78E−1 |
| H 4-1 | 1.75 | Ots13 | 2.05E−1 | 8.65E−2 | 2.86E−1 | 1.20E−1 |
| Hb 4 | 3.70 | Gar12 | 2.35E−2 | 2.41E−2 | 1.96E−1 | 2.01E−1 |
| Hen 2-118 | 1.70 | WL07 | 8.70E−2 | 5.58E−2 | 8.13E−2 | 5.21E−2 |
| Hen 2-155 | 6.30 | Jon15 | — | — | 1.41E−1 | — |
| Hen 2-158 | 1.61 | Gar18 | 3.73E−1 | 1.89E−1 | 2.80E−1 | 1.42E−1 |
| Hen 2-161 | 11.00 | Jon15 | — | — | 2.69E−1 | — |
| Hen 2-283 | 6.00 | Wes18 | — | — | 1.38E−1 | — |
| Hen 2-436 | 16.30 | Ots11 | — | — | 2.25E−2 | 4.04E−2 |
| Hen 2-73 | 2.29 | Gar18 | 1.48E−1 | 9.47E−2 | 1.81E−1 | 1.16E−1 |
| Hen 2-86 | 1.90 | Gar12 | 5.65E−2 | 4.94E−2 | 6.03E−3 | 5.27E−3 |
| Hen 2-96 | 1.41 | Gar18 | 1.05E−1 | 5.29E−2 | 1.33E−1 | 6.74E−2 |
| Hen 3-1357 (a) | 1.51 | Ots17 | 3.61E−2 | 3.39E−2 | — | — |
| Hen 3-1357 (b) | 2.52 | Pen22 | 5.83E−2 | — | — | — |
| Hen 3-1357 (c)* | 3.47 | Pen22 | 7.10E−2 | 3.31E−2 | — | — |
| Hf 2-2 (a) | 70.00 | Liu06 | — | — | — | — |
| Hf 2-2 (b)* | 80.00 | McN16 | — | — | 8.55E−2 | 1.73E−2 |
| Hf 2-2 (c) | 83.00 | Wes18 | — | — | — | — |
| Hu 1-1 | 2.97 | Wes05 | 5.15E−1 | — | 5.15E−1 | — |
| Hu 1-2 | 1.60 | Liu04A | 1.33E−1 | 1.08E−1 | 1.33E−1 | 1.08E−1 |
| Hu 2-1 (a)* | 1.85 | RP22 | — | 1.05E−1 | — | 1.05E−1 |
| Hu 2-1 (b) | 4.00 | Wes05 | — | — | — | — |
| IC 351 | 3.14 | Wes05 | 6.17E−2 | — | 6.17E−2 | — |
| IC 418 | 2.00 | Sha04 | — | — | 8.48E−2 | 1.19E−1 |
| IC 1747 | 3.20 | Wes05 | 6.86E−2 | — | 6.86E−2 | — |
| IC 2003 | 7.31 | Wes05 | 8.78E−2 | 2.08E−1 | 8.78E−2 | 2.08E−1 |
| IC 3568 | 2.20 | Liu04A | 1.45E−1 | — | 1.45E−1 | — |
| IC 4191 (a)* | 2.40 | Tsa04 | 5.52E−2 | 3.67E−2 | 5.42E−2 | 3.59E−2 |
| IC 4191 (b) | 2.40 | Tsa04 | 6.85E−1 | 4.49E−2 | 6.71E−1 | 4.40E−2 |
| IC 4406 | 1.90 | Tsa04 | 6.30E−2 | 1.40E−2 | 6.44E−2 | 1.44E−2 |
| IC 4593 | 3.60 | Rob05 | 4.13E−1 | 5.76E−1 | 4.13E−1 | 5.76E−1 |
| IC 4663 | 39.00 | Moh23 | — | — | 4.94E−2 | — |
| IC 4699 | 6.20 | WL07 | 5.46E−2 | — | 5.57E−2 | — |
| IC 4776 | 1.75 | Sow17 | — | — | 6.86E−2 | 4.25E−2 |
| IC 4846 (a)* | 1.50 | WL07 | 2.09E−1 | 2.22E−1 | 1.90E−1 | 2.02E−1 |
| IC 4846 (b) | 2.91 | Wes05 | 8.27E−2 | — | 8.27E−2 | — |
| IC 4997 | 4.87 | RP22 | — | — | 9.00E−2 | — |
| IC 5217 | 2.26 | Wes05 | — | — | — | — |
| K 648 | 2.99 | Ots15A | 5.29E−2 | 2.72E−2 | 5.36E−2 | 2.76E−2 |
| M 1-20 | 1.40 | WL07 | 2.40E−1 | 1.83E−1 | 2.47E−1 | 1.87E−1 |
| M 1-25 | 1.80 | Gar13 | 5.49E−2 | — | 1.31E−1 | — |
| M 1-29 | 3.00 | WL07 | 1.98E−1 | 1.14E−1 | 3.22E−1 | 1.85E−1 |
| M 1-30 | 2.40 | Gar13 | 3.24E−2 | 2.79E−2 | 1.06E−1 | 9.11E−2 |

**Table A4 – Continued**

| Name | ADF | Ref. ADF | $M_{ion}$([S II]) Slit ($M_\odot$) | $M_{ion}$([Cl III]) Slit ($M_\odot$) | $M_{ion}$([S II]) Global ($M_\odot$) | $M_{ion}$([Cl III]) Global ($M_\odot$) |
|---|---|---|---|---|---|---|
| M 1-31 | 2.11 | Gar18 | 3.07E−1 | 7.95E−2 | 1.010 | 2.61E−1 |
| M 1-32 | 3.00 | Gar13 | 4.71E−3 | 4.79E−3 | 1.48E−2 | 1.50E−2 |
| M 1-33 | 2.77 | Gar18 | 2.78E−1 | 1.94E−1 | 2.62E−1 | 1.83E−1 |
| M 1-42 (a)* | 20.00 | Liu01 | 2.54E−1 | 1.59E−1 | 2.60E−1 | 1.63E−1 |
| M 1-42 (b) | 22.00 | McN16 | 2.33E−1 | 1.66E−1 | 2.39E−1 | 1.70E−1 |
| M 1-60 | 2.75 | Gar18 | 2.46E−1 | 1.31E−1 | 2.07E−1 | 1.10E−1 |
| M 1-61 (a)* | 1.60 | Gar13 | 6.84E−2 | 8.53E−2 | 1.04E−1 | 1.29E−1 |
| M 1-61 (b) | 2.00 | WL07 | 7.80E−2 | 5.27E−2 | 8.36E−2 | 5.65E−2 |
| M 1-73 | 3.61 | Wes05 | 1.95E−1 | — | 1.95E−1 | — |
| M 1-74 | 2.14 | Wes05 | 1.84E−1 | — | 1.84E−1 | — |
| M 2-4 | 1.90 | WL07 | 2.93E−1 | 2.07E−1 | 3.69E−1 | 2.61E−1 |
| M 2-6 | 2.30 | WL07 | 7.62E−2 | 4.33E−1 | 1.63E−1 | 9.27E−1 |
| M 2-23 | 1.40 | WL07 | 1.13E−1 | 1.01E−1 | 1.16E−1 | 1.04E−1 |
| M 2-24 | 17.00 | Zha03 | 3.77E−1 | — | 3.86E−1 | — |
| M 2-27 | 2.20 | WL07 | 9.58E−2 | 4.11E−2 | 1.03E−1 | 4.40E−2 |
| M 2-31 | 2.42 | Gar18 | 1.62E−1 | 9.06E−2 | 1.79E−1 | 1.00E−1 |
| M 2-33 | 2.20 | WL07 | 2.75E−1 | — | 4.90E−1 | — |
| M 2-36 (a) | 5.00 | Liu01 | 9.09E−2 | 5.60E−2 | 1.62E−1 | 9.98E−2 |
| M 2-36 (b)* | 6.76 | Esp21 | 1.18E−1 | 7.27E−2 | 2.11E−1 | 1.30E−1 |
| M 2-39 | 0.40 | WL07 | 7.99E−2 | 1.99E−1 | 6.95E−2 | 1.73E−1 |
| M 2-42 | 2.10 | WL07 | 2.17E−1 | 1.85E−1 | 2.08E−1 | 1.77E−1 |
| M 3-7 | 4.40 | WL07 | 1.48E−1 | 2.27E−1 | 1.82E−1 | 2.79E−1 |
| M 3-15 | 2.30 | Gar13 | 6.20E−1 | 4.46E−1 | 2.44E−1 | 1.76E−1 |
| M 3-21 | 2.60 | WL07 | 7.45E−2 | 5.57E−2 | 6.96E−2 | 5.20E−2 |
| M 3-27 (a) | 5.48 | Wes05 | — | — | — | — |
| M 3-27 (b)* | 3.26 | RPB24 | — | — | 9.61E−1 | — |
| M 3-29 | 2.20 | WL07 | 1.60E−1 | — | 2.41E−1 | — |
| M 3-32 | 17.80 | WL07 | 1.48E−1 | 3.53E−1 | 1.33E−1 | 3.16E−1 |
| M 3-33 | 6.60 | WL07 | 2.41E−1 | 1.32E−1 | 2.06E−1 | 1.13E−1 |
| M 3-34 | 4.23 | Wes05 | — | — | — | — |
| Me 2-2 | 2.10 | Wes05 | 1.83 | — | 1.83 | — |
| MPA J1759-3007 | 62.00 | Wes18 | — | — | — | — |
| MyCn 18 | 1.80 | Tsa04 | 1.05E−1 | 4.20E−2 | 1.26E−1 | 5.05E−2 |
| NGC 40 | 17.80 | Liu04A | 6.23E−1 | 1.07 | 3.19E−1 | 5.50E−1 |
| NGC 1501 | 32.00 | Erc04 | — | — | 1.16E−1 | 5.18E−2 |
| NGC 2022 | 16.00 | Tsa04 | 1.81E−1 | 8.28E−2 | 1.81E−1 | 8.28E−2 |
| NGC 2392 (a) | 1.65 | Zha12 | — | — | 1.31E−1 | 2.09E−1 |
| NGC 2392 (b)* | 1.65 | Zha12 | — | — | 4.76E−1 | 2.97E−1 |
| NGC 2440 | 5.40 | Tsa04 | 3.32E−2 | — | 3.32E−2 | — |
| NGC 2867 (a)* | 1.49 | Gar09 | — | — | 1.27E−1 | 1.02E−1 |
| NGC 2867 (b) | 1.77 | Gar09 | — | — | 1.96E−1 | 1.09E−1 |
| NGC 3132 | 2.40 | Tsa04 | 2.71E−1 | 1.65E−1 | 2.71E−1 | 1.65E−1 |
| NGC 3242 | 2.20 | Tsa04 | 3.20E−1 | 4.05E−1 | 3.20E−1 | 4.05E−1 |
| NGC 3918 (a) | 1.80 | Tsa04 | 1.62E−1 | 1.19E−1 | 1.62E−1 | 1.19E−1 |
| NGC 3918 (b)* | 1.85 | Gar15 | 5.84E−3 | 4.86E−3 | 9.61E−2 | 8.00E−2 |
| NGC 5189 | 1.60 | Gar13 | 1.81E−4 | 1.59E−4 | 4.67E−2 | 4.10E−2 |
| NGC 5307 | 1.90 | Rui03 | 1.01E−2 | 2.82E−2 | 5.64E−2 | 1.58E−1 |
| NGC 5315 (a)* | 1.58 | Ma17 | — | — | 8.96E−2 | 2.85E−2 |
| NGC 5315 (b) | 2.00 | Tsa04 | 1.13E−1 | 2.88E−2 | 1.13E−1 | 2.88E−2 |
| NGC 5882 | 2.10 | Tsa04 | 1.95E−1 | 2.06E−1 | 1.99E−1 | 2.10E−1 |
| NGC 6153 (a)* | 10.00 | Liu00 | — | — | 2.30E−1 | 1.83E−1 |
| NGC 6153 (b) | 11.00 | McN16 | — | — | 2.28E−1 | — |
| NGC 6210 (a) | 2.50 | Rob05 | — | — | — | — |
| NGC 6210 (b)* | 3.10 | Liu04A | 1.92E−1 | 1.91E−1 | 1.92E−1 | 1.91E−1 |
| NGC 6302 | 3.60 | Tsa04 | 4.41E−2 | 1.67E−2 | 4.41E−2 | 1.67E−2 |
| NGC 6326 | 23.00 | Wes18 | — | — | 4.06E−1 | — |
| NGC 6337 | 18.00 | Wes18 | — | — | 1.59E−1 | — |
| NGC 6369 | 1.40 | Gar13 | 2.63E−3 | 2.44E−3 | 1.88E−1 | 1.74E−1 |
| NGC 6439 | 6.20 | WL07 | 2.74E−1 | 2.12E−1 | 3.08E−1 | 2.38E−1 |
| NGC 6543 (a) | 2.07 | Rob05 | — | — | — | — |
| NGC 6543 (b) | 2.74 | Rob05 | — | — | — | — |
| NGC 6543S (c)* | 3.00 | Wes04 | — | — | 1.38E−1 | 1.44E−1 |
| NGC 6565 | 1.70 | WL07 | 1.44E−1 | 1.28E−1 | 1.55E−1 | 1.37E−1 |
| NGC 6567 | 2.20 | WL07 | 1.71E−1 | 6.24E−2 | 1.74E−1 | 6.35E−2 |

Table A4 – Continued

| Name | ADF | Ref. ADF | $M_{ion}$([S II]) Slit ($M_\odot$) | $M_{ion}$([Cl III]) Slit ($M_\odot$) | $M_{ion}$([S II]) Global ($M_\odot$) | $M_{ion}$([Cl III]) Global ($M_\odot$) |
|---|---|---|---|---|---|---|
| NGC 6572 (a) | 1.38 | Rob05 | — | — | — | — |
| NGC 6572 (b) | 1.50 | Tsa04 | — | — | — | — |
| NGC 6572 (c) | 1.54 | Rob05 | — | — | — | — |
| NGC 6572 (d)* | 1.60 | Liu04A | 1.26E−1 | 4.74E−1 | 1.26E−1 | 4.74E−1 |
| NGC 6620 | 3.20 | WL07 | 1.35E−1 | 9.99E−2 | 1.35E−1 | 9.99E−2 |
| NGC 6644 | 1.90 | Tsa04 | — | — | — | — |
| NGC 6720 (a)* | 2.40 | Liu04A | 2.25E−1 | 1.86E−1 | 2.25E−1 | 1.86E−1 |
| NGC 6720 (b) | 5.00 | Gar01 | — | — | — | — |
| NGC 6741 | 1.90 | Liu04A | 2.06E−1 | 1.67E−1 | 2.15E−1 | 1.75E−1 |
| NGC 6778 | 18.00 | Jon16 | 2.67E−1 | — | 2.99E−1 | — |
| NGC 6790 (a) | 1.44 | Rob05 | — | — | — | — |
| NGC 6790 (b)* | 1.70 | Liu04A | 1.01E−1 | 1.36E−1 | 1.01E−1 | 1.36E−1 |
| NGC 6803 | 2.71 | Wes05 | 2.35E−1 | 1.51E−1 | 2.35E−1 | 1.51E−1 |
| NGC 6807 | 2.00 | Wes05 | 1.81E−1 | — | 1.81E−1 | — |
| NGC 6818 | 2.90 | Tsa04 | 2.25E−1 | 1.36E−1 | 2.25E−1 | 1.36E−1 |
| NGC 6826 | 1.90 | Liu04A | 1.61E−1 | 1.84E−1 | 1.54E−1 | 1.76E−1 |
| NGC 6833 | 2.47 | Wes05 | — | — | — | — |
| NGC 6879 | 2.46 | Wes05 | — | — | — | — |
| NGC 6884 | 2.30 | Liu04A | 1.58E−1 | 1.76E−1 | 1.58E−1 | 1.76E−1 |
| NGC 6891 | 1.52 | Wes05 | — | — | — | — |
| NGC 7009 (a) | 5.00 | Liu95 | — | — | — | — |
| NGC 7009 (b)* | 5.00 | Fan11 | — | — | 1.50E−1 | 1.39E−1 |
| NGC 7026 | 3.36 | Wes05 | — | 2.48E−1 | — | 2.48E−1 |
| NGC 7027 (a) | 3.58 | Rob05 | — | — | — | — |
| NGC 7027 (b)* | 1.30 | Zha05 | — | — | 2.65E−2 | 2.54E−2 |
| NGC 7662 | 2.00 | Liu04A | 2.90E−1 | 3.17E−1 | 2.90E−1 | 3.17E−1 |
| Ou 5 | 56.00 | Cor15 | 2.43E−2 | — | — | — |
| PB 8 | 2.57 | Gar09 | — | — | 1.29E−1 | 1.44E−1 |
| PC 14 | 1.90 | Gar13 | 1.97E−2 | 2.00E−2 | 7.25E−2 | 7.37E−2 |
| Pe 1-1 | 1.70 | Gar13 | 3.76E−2 | 2.19E−2 | 8.06E−2 | 4.69E−2 |
| Pe 1-9 | 60.00 | Wes18 | — | — | 8.09E−2 | — |
| Sp 3 | 24.00 | Mis19 | — | — | 6.29E−2 | 3.96E−2 |
| Sp 4-1 | 2.94 | Wes05 | — | — | — | — |
| Vy 1-2 (a)* | 5.34 | RP22 | 9.17E−2 | 4.87E−2 | 9.17E−2 | 4.87E−2 |
| Vy 1-2 (b) | 6.17 | Wes05 | 3.48E−1 | — | 3.48E−1 | — |
| Vy 2-1 | 2.00 | WL07 | 3.34E−1 | 1.76E−1 | 3.92E−1 | 2.06E−1 |
| Vy 2-2 (a)* | 4.30 | RP22 | 3.82E−2 | 3.43E−3 | 3.82E−2 | 3.43E−3 |
| Vy 2-2 (b) | 11.80 | Wes05 | — | — | — | — |
| Wray 16-423 | 1.64 | Ots15B | 2.20E−1 | 1.23E−1 | 1.44E−1 | 8.06E−2 |

* Adopted mass and ADF values

References: Cor15: Corradi et al. (2015), Erc04: Ercolano et al. (2004), Esp21: Espíritu & Peimbert (2021),
Fan11: Fang & Liu (2011), Gar09: García-Rojas et al. (2009), Gar13: García-Rojas et al. (2013),
Gar15: García-Rojas et al. (2015), Gar18: García-Rojas et al. (2018), Jon15: Jones et al. (2015),
Jon16: Jones et al. (2016), Liu00: Liu et al. (2000), Liu04A: Liu et al. (2004a), Liu06: Liu et al. (2006),
Ma17: Madonna et al. (2017), McN16: McNabb et al. (2016), Mis19: Miszalski et al. (2019),
Moh23: Mohery et al. (2023), Ots09: Otsuka et al. (2009), Ots10: Otsuka et al. (2010), Ots11: Otsuka et al. (2011),
Ots13: Otsuka & Tajitsu (2013), Ots15A: Otsuka et al. (2015), Ots15B: Otsuka (2015), Ots17: Otsuka et al. (2017),
Pen22: Peña et al. (2022), Rob05: Robertson-Tessi & Garnett (2005), Rui03: Ruiz et al. (2003),
RP22: Ruiz-Escobedo & Peña (2022), RPB24: Ruiz-Escobedo et al. (2024), Sha04: Sharpee et al. (2004),
Sim22: Simpson et al. (2022), Sow17: Sowicka et al. (2017), Tsa04: Tsamis et al. (2004), WL07: Wang & Liu (2007),
Wes03: Wesson et al. (2003), Wes04: Wesson & Liu (2004), Wes05: Wesson et al. (2005), Wes18: Wesson et al. (2018),
Zha05: Zhang et al. (2005), Zha12: Zhang et al. (2012).

This paper has been typeset from a T<sub>E</sub>X/LaT<sub>E</sub>X file prepared by the author.